\documentclass[twocolumn]{autart}    

\usepackage{amssymb,latexsym,amsfonts,amsmath}

\usepackage{wrapfig} 

\newenvironment{authorbio}[2][]{%
	\par\addvspace{1em}%
	\noindent
	\if\relax\detokenize{#1}\relax 
	\else
	\begin{wrapfigure}{l}{0.8in} 
		\vspace{-5pt} 
		\includegraphics[width=1in,height=1.25in,clip,keepaspectratio]{#1}
	\end{wrapfigure}%
	\fi
	\noindent\textbf{#2} 
}{\par\addvspace{2em}}

\usepackage{mathrsfs}

\usepackage{algorithm}
\usepackage{algcompatible}

\usepackage{natbib}

\usepackage{xspace}

\usepackage[nocomma]{optidef}

\usepackage{graphicx}
\usepackage{paralist}
\usepackage{dsfont}
\usepackage{caption} 
\usepackage{booktabs}
\usepackage{subfig}
\usepackage{eso-pic}
\usepackage{epsfig}
\usepackage{mathtools}
\usepackage{xfrac}
\usepackage {url}
\usepackage{epstopdf}
\usepackage{tikz}
\usepackage{IEEEtrantools}
\usetikzlibrary{arrows,calc,decorations.markings,math,arrows.meta}
\newtheorem{theorem}{Theorem}[section]
\newtheorem{lemma}[theorem]{Lemma}

\newtheorem{definition}[theorem]{Definition}
\newtheorem{example}[theorem]{Example}

\newtheorem{remark}[theorem]{Remark}
\newtheorem{assumption}[theorem]{Assumption}
\numberwithin{equation}{section}

\usepackage[many]{tcolorbox}
\usetikzlibrary{calc}
\tcbuselibrary{skins}

\newtcolorbox{resp}[1][]{%
	enhanced jigsaw,%
	colback=gray!5!white,%
	colframe=gray!80!black,%
	size=small,%
	boxrule=1pt,%
	halign title=flush center,%
	coltitle=black,%
	breakable,%
	drop shadow=black!50!white,%
	attach boxed title to top left={xshift=1cm,yshift=-\tcboxedtitleheight/2,yshifttext=-\tcboxedtitleheight/2},%
	minipage boxed title=3cm,%
	boxed title style={%
		colback=white,%
		size=fbox,%
		boxrule=1pt,%
		boxsep=2pt,%
		underlay={%
			\coordinate (dotA) at ($(interior.west) + (-0.5pt,0)$);
			\coordinate (dotB) at ($(interior.east) + (0.5pt,0)$);
			\begin{scope}[gray!80!black]
				\fill (dotA) circle (2pt);
				\fill (dotB) circle (2pt);
			\end{scope}
		}%
	},%
	#1%
}

\definecolor{mydarkred}{HTML}{990000}

\newcommand{\circnum}[1]{%
	\tikz[baseline=(char.base)]{
		\node[
		circle,
		draw=mydarkred,
		fill=white,
		line width=1pt,
		inner sep=0pt,
		minimum size=1em,
		text height=1.5ex,
		text depth=.25ex,
		font=\fontsize{7.5}{7.5}\selectfont
		] (char) {#1};
	}%
}

\newcommand{\R}{{\mathbb{R}}}

\newcommand{\ie}{{\it i.e.}}

\newcommand{\argmin}{\textrm{arg}\min}

\makeatletter
\long\def\@maketablecaption#1#2{\@tablecaptionsize
	\global \@minipagefalse
\hbox to \hsize{\parbox[t]{\hsize}{\centering #1 \\ #2}}}

\makeatother
\allowdisplaybreaks
\begin{document}

\begin{frontmatter}
	
	\title{Data-Driven Incremental GAS Certificate of Nonlinear Homogeneous Networks: {A Scenario Approach with Noisy Data}}


\author[UK1]{Mahdieh Zaker}\ead{m.zaker2@newcastle.ac.uk},
\author[{UK2,Italy}]{David Angeli}\ead{d.angeli@imperial.ac.uk}, \author[UK1]{Abolfazl Lavaei}\ead{abolfazl.lavaei@newcastle.ac.uk} 

\address[UK1]{School of Computing, Newcastle University, United Kingdom}                   
\address[UK2]{Department of Electrical and Electronic Engineering, Imperial College London, United Kingdom}
\address[Italy]{Department of Information Engineering, University of Florence, Italy}


\begin{abstract}   
	This work focuses on a \emph{compositional data-driven} approach to verify \emph{incremental global asymptotic stability} ($\delta$-GAS) over interconnected homogeneous networks of degree one with unknown mathematical dynamics. Our proposed approach leverages the concept of incremental input-to-state stability ($\delta$-ISS) of subsystems, characterized by $\delta$-ISS Lyapunov functions. To implement our data-driven scheme, we initially reframe the $\delta$-ISS Lyapunov conditions as a robust optimization program (ROP). Due to the presence of unknown subsystem dynamics in the ROP constraints, we develop a scenario optimization program (SOP) by gathering data from trajectories of each unknown subsystem. {However, since the measured one-step transition data are corrupted by noise with a known bound on its norm, rendering the proposed SOP intractable, we introduce an auxiliary SOP that explicitly accommodates noisy measurements. We solve the auxiliary SOP} and construct a $\delta$-ISS Lyapunov function for each subsystem with unknown dynamics. We then leverage a small-gain compositional condition to facilitate the construction of an \emph{incremental Lyapunov function} for an unknown interconnected network based on the data-driven $\delta$-ISS Lyapunov functions of its individual subsystems, while providing correctness guarantees, incorporating the bound on the noise norm. We demonstrate that our data-driven compositional approach {reduces the sample complexity to the subsystem level.} In contrast, the existing monolithic approach in the literature exhibits \emph{exponential} growth in sample complexity {with the total network size, rendering it impractical for real-world applications as the number of subsystems increases.} To validate the effectiveness of our compositional data-driven approach, we apply it to an unknown {controlled physical} nonlinear homogeneous network of degree one, comprising $10000$ subsystems. By gathering {noisy} data from each unknown subsystem, we demonstrate that the interconnected network is $\delta$-GAS with a correctness guarantee.
\end{abstract}

\end{frontmatter}

\section{Introduction}

Incremental input-to-state stability ($\delta$-ISS)~\citep{angeli,biemond2018incremental,van2023small,d2023incremental} is of paramount importance in the field of control and stability analysis due to its ability to provide robust guarantees in systems where traditional methods may be inadequate. This concept addresses the incremental behavior of a system in response to changes in inputs, offering insights into its stability properties over time. In particular, $\delta$-ISS centers on the convergence of trajectories relative to each other, rather than relative to a fixed trajectory or equilibrium point. By focusing on how small changes in inputs affect the system's state evolution, $\delta$-ISS enables the characterization of system behavior in the presence of uncertainties or variations in operating conditions. This capability is particularly valuable in real-world applications involving the analysis of complex nonlinear systems. Examples of such applications encompass synchronizing interconnected oscillators~\citep{stan2007analysis}, constructing symbolic models~\citep{pola2008approximately}, modeling nonlinear analog circuits~\citep{bond2010compact}, and synchronizing cyclic feedback systems~\citep{hamadeh2011global}, to name a few. 

Most stability analyses of dynamical systems in the literature rely on explicit models. Nonetheless, the principal obstacle arises from the absence of closed-form mathematical models for numerous complex and diverse systems, rendering model-based techniques impractical for analysis. Consequently, recent years have witnessed a significant surge in interest in \emph{data-driven} control methods due to their widespread applicability in various real-world systems, including but not limited to autonomous vehicles, (air) traffic networks, smart grids, and biological networks.  In this context, \emph{indirect} data-driven methodologies have been introduced, relying on \emph{identification techniques} to analyze unknown systems by learning their approximate models (see \emph{e.g.,}~\cite{Hou2013model}). However, acquiring accurate models for complex systems presents significant challenges and is often prohibitively expensive in practice. Moreover, even if the model can be identified using system identification techniques, the search for a valid Lyapunov function to demonstrate stability properties remains necessary. Consequently, computational complexity arises at both stages: model identification and Lyapunov function construction. Hence, the proposition of \emph{direct} data-driven approaches becomes crucial, circumventing the need for the system identification phase and directly leveraging system measurements to conduct the requisite analysis~\citep{dorfler2022bridging}.

{\bf State of the art.} In the last couple of decades, several findings have emerged in the domain of data-driven optimization methods, with noteworthy results proposed. Existing contributions include the scenario approach for robust control analysis and synthesis problems using semidefinite convex programming~\citep{calafiore2006scenario};
a framework for random convex programs to derive explicit upper tail probability bounds and extend to cases with a posteriori violated constraints~\citep{calafiore2010random}; a probabilistic connection between scenario convex programs and robust convex and chance-constrained programs, expanded to non-convex problems with binary decision variables~\citep{esfahani2014performance,margellos2014road}; and a constructive framework linking infinite-dimensional linear programming with tractable finite convex programs~\citep{mohajerin2018infinite}, among others. {Related finite-analysis ideas based on Lipschitz continuity have also been used to handle continuum constraints in system identification; see, for instance,~\citep[Sec.~5]{khosravi2023kernel}, where frequency-domain side-information is enforced on a finite grid via a Lipschitz-based mesh condition.}

More recently, emerging research has focused on the formal analysis of unknown dynamical systems through \emph{data-driven} approaches. Several notable contributions include a data-driven stability analysis of black-box switched linear systems using a constructed \emph{common} Lyapunov function~\citep{kenanian2019data}; a data-driven approach for the stability analysis of continuous-time unknown systems~\citep{boffi2020learning}; a data-driven technique for input-to-state stabilization~\citep{chen2025data}; a data-enabled predictive control algorithm for unknown stochastic linear systems~\citep{coulson2019regularized}; a data-driven analysis of discrete-time homogeneous systems~\citep{lavaei2022data}; {data-driven nonlinear system analysis using polynomial approximation and sum-of-squares techniques~\citep{martin2023data};} and guaranteeing stability for unknown nonlinear systems by simultaneously synthesizing a nonlinear controller and a neural Lyapunov function, with formal verification performed via Satisfiability Modulo Theories solvers~\citep{zhou2022neural}. {Recent neural-network-based and contraction-based methods have further expanded this landscape, including the synthesis of incrementally stabilizing neural controllers for unknown systems~\citep{basu2026neural}, distributed learning of incremental Lyapunov certificates for interconnected systems~\citep{basu2025scalable}, and learning-based control contraction metric methods, rooted in contraction theory~\citep{lohmiller1998contraction}, for tracking control with convergence or tracking-error bound guarantees~\citep{sun2021learning}.}

{While such studies offer promising insights, most existing approaches either provide \emph{local} guarantees only over compact sets or focus on monolithic systems, which limits their scalability to high-dimensional interconnected networks. To address these key limitations, we introduce an innovative data-driven, \emph{divide-and-conquer} strategy for {$\delta$-ISS} analysis in large-scale interconnected systems with {unknown} models and \emph{noisy measurements}, offering, to the best of our knowledge, the first such approach in this setting. Unlike previous approaches, which focus primarily on stability analysis or controller synthesis for monolithic systems, our method addresses the limitations of sample complexity associated with high-dimensional systems. Specifically, the number of required data samples in scenario-based stability certification exponentially increases with the state-space dimension in existing methods. However, our compositional approach significantly mitigates this sample complexity by reducing it to the level of subsystems, resulting in a \emph{manageable growth} in the required data as the number of subsystems increases. Moreover, while guarantees in prior works such as those in  \citep{kenanian2019data,rubbens2021data,boffi2020learning,basu2025scalable,basu2026neural} are probabilistic and/or local, our proposed approach offers a correctness guarantee with confidence level~$1$ (when the Lipschitz constant of the dynamics is given) and, under the degree-one homogeneity assumption, yields \emph{global} stability guarantees by lifting results from the unit sphere to the entire Euclidean space.}

\begin{figure*}[t!]
\centering
\includegraphics[width=\linewidth]{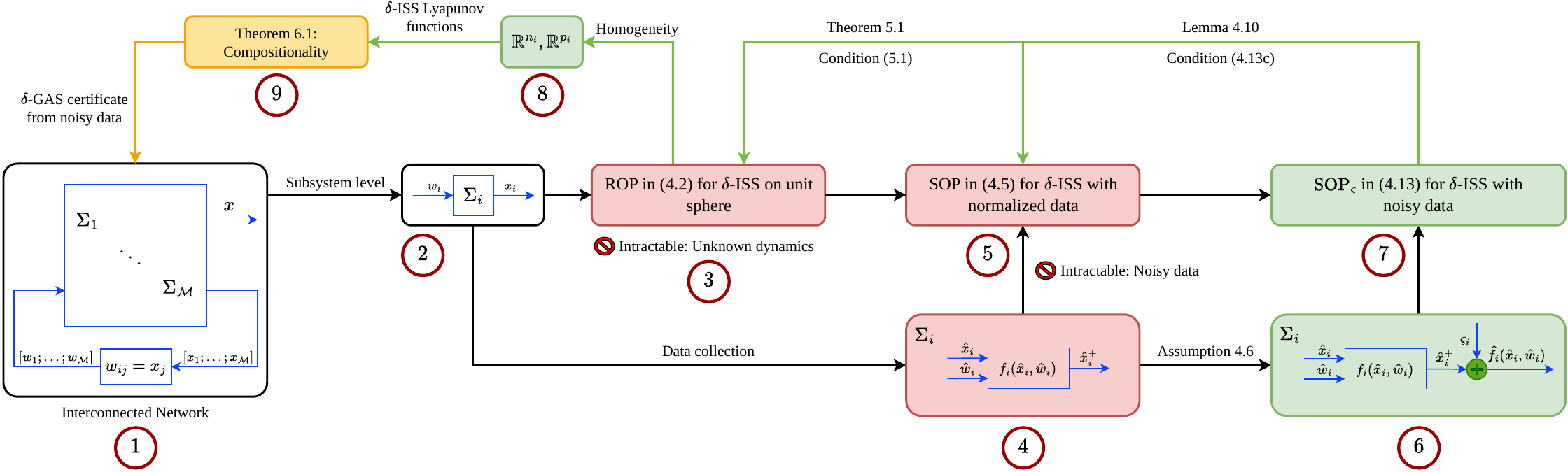}
\caption{{A thorough overview of the proposed framework with noisy data.}}
\label{fig:framework}
\end{figure*}

The literature on data-driven $\delta$-ISS analysis of unknown systems remains very limited. In this context, \cite{sundarsingh2024backstepping} employ an \emph{indirect} method based on Gaussian process regression to approximate the unknown dynamics of a particular class of control systems. A backstepping controller is then synthesized to achieve incremental input-to-state \emph{practical} stability, a relaxed variant of $\delta$-ISS. In addition, \cite{zaker2024certified} propose a \emph{direct} data-driven strategy that bypasses system identification altogether. Instead, it constructs a control law ensuring $\delta$-ISS by leveraging sufficiently exciting trajectory data. {More recently, \cite{basu2026neural} develop a neural-network-based framework for synthesizing incrementally input-to-state stabilizing controllers for unknown continuous-time systems through a learned $\delta$-ISS control Lyapunov function. While these studies offer valuable insights, they are limited to low-dimensional, monolithic systems and their certification steps are typically carried out over compact sets, yielding \emph{local} stability guarantees. Furthermore, our methodology fundamentally differs from these works by employing a scenario-based compositional framework with noisy data, which provides a distinct data-driven solution for $\delta$-ISS analysis in large-scale interconnected networks.}

{Some recent studies have aimed to address stability analysis of interconnected networks in data-driven settings~\citep{lavaei2023data,zaker2025data,zaker2026data,basu2025scalable}. However, the resulting guarantees and certificate structures differ remarkably from the ones presented in our work. In particular, \citep{lavaei2023data,zaker2025data} establish stability by leveraging the ISS properties of individual subsystems integrated with small-gain conditions, whereas \cite{basu2025scalable} learns subsystem-level neural certificates for \emph{local} $\delta$-ISS verification in interconnected systems, where each certificate depends on the state of a subsystem and those of its neighbors. In contrast, our work is the first to investigate \emph{global} incremental GAS certificates for interconnected networks in a direct data-driven setting by integrating decentralized subsystem-level $\delta$-ISS Lyapunov functions, each depending only on the local state, with a compositional small-gain condition. In addition, none of the aforementioned studies consider measurement noise, as we do in this work, which makes the development significantly more challenging but instead enhances its practical relevance.} 

{\textbf{Original contribution.} The primary contribution of this work is the development of a compositional data-driven framework for certifying incremental global asymptotic stability ($\delta$-GAS) from noisy data. Fig.~\ref{fig:framework} summarizes the high-level structure of the proposed approach. As illustrated, we initially consider an interconnected network of degree-one homogeneous subsystems with unknown dynamics (Step \circnum{$1$}). Our method exploits subsystem-level incremental input-to-state stability, characterized by suitable $\delta$-ISS Lyapunov functions (Step \circnum{$2$}). To render the approach data-driven, we first recast the $\delta$-ISS Lyapunov conditions as a robust optimization program (ROP) posed over the compact unit-sphere set (Step \circnum{$3$}). Since the constraints of this ROP depend on unknown dynamics, we then formulate a scenario optimization program (SOP) using noise-free data over the unit sphere collected from each unknown subsystem (Steps \circnum{$4$}, \!\circnum{$5$}). However, since the one-step-ahead data are corrupted by measurement noise (Step \circnum{$6$}), the SOP associated with noise-free data cannot be solved directly. We therefore introduce an auxiliary SOP with new conditions that constructs $\delta$-ISS Lyapunov functions for the individual subsystems with unknown dynamics directly from noisy data (Step \circnum{$7$}). After solving the auxiliary SOP and establishing the validity of the obtained solution, first for the original SOP and, consequently, for the ROP, we exploit the homogeneity of both the subsystem dynamics and the $\delta$-ISS Lyapunov functions to extend the results from the compact unit-sphere set to the entire Euclidean space (Step \circnum{$8$}). We then derive a compositional condition, based on small-gain reasoning, to construct an incremental Lyapunov function for the unknown interconnected network based on  $\delta$-ISS Lyapunov functions designed from noisy data, while providing correctness guarantees (Step \circnum{$9$}).}

We showcase how our data-driven compositional strategy circumvents the sample complexity issue inherent in monolithic-based data-driven methodologies aimed at certifying properties over unknown systems using data. {In particular, our approach decomposes the sample complexity to the subsystem level, where the required number of samples scales exponentially only with the effective dimensions at the \emph{subsystem level}, rather than the total network size.} To assess the efficacy of our data-driven approach, we apply it to {a physical case study, namely a network of controlled Duffing oscillators,} encompassing $10000$ subsystems with unknown dynamics.

\section{Problem Description}\label{Problem_Description}

\subsection{Notation and Preliminaries}
Sets of real, non-negative and positive real numbers are denoted by $\mathbb{R}$, $\mathbb{R}^+_0$, and $\mathbb{R}^+$, respectively. We denote sets of non-negative and positive integers by $\mathbb{N} \coloneq \{0,1,2,\ldots\}$ and $\mathbb{N}^+=\{1,2,\ldots\}$, respectively. Given $N$ vectors $x_i \in \mathbb{R}^{n_i}$, $x=[x_1;\ldots;x_N]$ denotes the corresponding column vector of dimension $\sum_{i=1}^{N} n_i$. 
We represent the Euclidean norm of a vector $x\in\mathbb{R}^{n}$ by $\Vert x\Vert$. Given sets $X_i, i\in\{1,\ldots,N\}$,
their Cartesian product is denoted by $\prod_{i=1}^{N}X_i$. For a symmetric matrix $P$, we denote its minimum and maximum eigenvalues as $\lambda_{\min}(P)$ and $\lambda_{\max}(P)$, respectively.
Constructing matrix $A$ with $a_{ij}$ element on its $i$-th row and $j$-th column is represented by $A=\{a_{ij}\}$. A column vector in $\R^{n}$ with all elements equal to one is represented by $\mathds{1}_n$.  {For a differentiable function \(f:\R^n \to \R\), \(\nabla f(x)\) denotes the gradient of \(f\) at \(x\), \ie, $\nabla f(x) :=\big[\frac{\partial f}{\partial x_1}(x);\dots;\frac{\partial f}{\partial x_n}(x)\big]$.} A function $\beta\!: \mathbb R_{0}^+\rightarrow \mathbb R_{0}^+,$ is said to be a class $\mathcal{K}$ function if it is continuous, strictly increasing, and $\beta(0)=0$. {A function $\beta\in\mathcal K$ belongs to class $\mathcal K_\infty$ if $\beta(r)\to\infty$ as $r\to\infty$.} A function $\beta: \mathbb{R}_{0}^+ \times \mathbb{R}_{0}^+ \rightarrow \mathbb{R}_{0}^+$ is said to belong to class $\mathcal {KL}$ if, for each fixed $s$, the map
$\beta(r,s)$ belongs to class $\mathcal K$ with respect to $r$ and, for each fixed $r>0$, the map $\beta(r,s)$ is decreasing with respect to $s$, and
$\beta(r,s)\rightarrow 0$ as $s \rightarrow \infty$.

\subsection{Discrete-Time Nonlinear Homogeneous Networks}

In this work, we investigate interconnected networks comprising discrete-time nonlinear homogeneous subsystems of degree one, as articulated in the subsequent definition.

\begin{definition}
A discrete-time nonlinear subsystem (dt-NS) is described by the tuple
\begin{equation}\label{Eq:1}
	\Sigma_i = (\mathbb R^{n_i},\mathbb R^{p_i},f_i),
\end{equation}
where:
\begin{itemize}
	\item $\mathbb R^{n_i}$ is the state set of dt-NS;
	\item $\mathbb R^{p_i}$ is the internal input set of dt-NS;
	\item $f_i:\mathbb R^{n_i}\times \mathbb R^{p_i}\rightarrow \mathbb R^{n_i}$ denotes a continuous function governing the system's evolution. {We make the assumption that $f_i$ is a homogeneous function of degree one, i.e., for any $\eta > 0$ and $x_i\in \mathbb R^{n_i},w_i\in \mathbb R^{p_i}$, $f_i(\eta x_i, \eta w_i) = \eta f_i(x_i,w_i)$~\citep[Definition 1.1]{polyakov2020generalized}.} We presume the map $f_i$ remains unknown to us.
\end{itemize}

Given an initial state $x_i(0)\in \mathbb R^{n_i}$ and an internal input sequence $w_i(\cdot)\!: \mathbb N \rightarrow \mathbb R^{p_i}$, the evolution of dt-NS $\Sigma_i$ can be characterized as
\begin{align}\label{Eq:2}
	\Sigma_i\!:{x}_i(k+1)=f_i(x_i(k),w_i(k)),\quad \quad k\in\mathbb N.
\end{align}
\end{definition}
With the primary aim of establishing an \emph{incremental} stability certificate for an interconnected dt-NS, we introduce a formal definition of such interconnected networks that does not involve the internal input $w_i$.

\begin{definition}\label{Def:3}
Let us consider $\mathcal M\!\in\!\mathbb N^+$ dt-NS $\Sigma_i\!=\!(\mathbb R^{n_i},\mathbb R^{p_i},f_i)$, where $i\in \{1,\dots,\mathcal M\}$.
The interconnection of $(\Sigma_i)_{i=1}^{\mathcal M}$ is denoted as
$\Sigma=(\mathbb R^{n},f)$, represented by
$\mathcal{I}(\Sigma_1,\ldots,\Sigma_{\mathcal M})$, where $n = \sum_{i = 1}^{\mathcal M} n_i$, and $f(x)\coloneq[f_1(x_1,w_1);\dots;f_{\mathcal M}(x_{\mathcal M},w_{\mathcal M})]$, after applying the following interconnection constraint:
\begin{align}\label{Eq:41}
	\forall i,j\in \{1,\dots,\mathcal M\},i\neq j\!: ~w_{ij}=x_{j},
\end{align}
where $w_{ij}$ are elements of the internal inputs partitioned as follows:
\begin{align}\label{Eq:31}
	w_i&=[{w_{i1};\ldots;w_{i(i-1)};w_{i(i+1)};\ldots;w_{i\mathcal M}}].
\end{align} 
The resulting interconnected dt-NS can be characterized as
\begin{equation}\label{Eq:51}
	\Sigma\!:{x}(k+1)=f(x(k)), \quad \text{with}~ f: \mathbb R^{n} \rightarrow \mathbb R^{n}.
\end{equation}
\end{definition}
{It is evident that due to the degree-one homogeneity of all $\Sigma_i, i\in\{1,\dots,\mathcal M\}$, the interconnected dt-NS $\Sigma$ also preserves its homogeneity with the same degree.} In other words, for any $\eta > 0$ and $x\in \mathbb R^{n}$, it holds that 
\begin{align*}
f(\eta x) &= [f_1(\eta x_1,\eta w_1);\dots;f_{\mathcal M}(\eta x_{\mathcal M},\eta w_{\mathcal M})] \\
&= [\eta f_1(x_1,w_1);\dots;\eta f_{\mathcal M}(x_{\mathcal M},w_{\mathcal M})] \\
&=\eta [f_1(x_1,w_1);\dots;f_{\mathcal M}(x_{\mathcal M},w_{\mathcal M})] = \eta f(x).
\end{align*}
{\begin{remark}\label{rem:hom}
	When the network dynamics are unknown and the measurements are noise-free, degree-one homogeneity can readily be examined empirically by performing experiments from several nonzero proportional initial conditions. More precisely, consider $N$ experiments with initial conditions $x_{0,l}$, $l \in \{1,\ldots,N\}$, chosen such that $x_{0,i}=\eta_{ij}x_{0,j}$ for all $i \neq j$. If the measured one-step transitions satisfy $f(x_{0,i})=\eta_{ij}f(x_{0,j})$ for all $i,j$, {i.e.}, if the proportional relation between the initial conditions is preserved by the corresponding one-step transitions, then this provides empirical evidence that the interconnected network is homogeneous of degree one.
	
	In case of noisy measurements, when the measured one-step transition $f(x)$ is corrupted by an unknown disturbance with known norm bound, the same idea can be adapted by repeating each experiment over $T$ trials. Specifically, let the empirical mean of the one-step transitions be
	$$
	\tilde f(x_{0,i})=\frac{1}{T}\sum_{z=1}^{T} f_z(x_{0,i}),
	$$
	where $f_z(\cdot)$ denotes the $z$-th noisy measurement. If the empirical means satisfy $\tilde f(x_{0,i})\approx\tilde \eta_{ij}\tilde f(x_{0,j})$ for all $i,j$, and if the deviation between $\tilde \eta_{ij}$ and $\eta_{ij}$ is sufficiently small relative to the given bound, then this provides empirical evidence, up to the measurement uncertainty, that the interconnected network is homogeneous of degree one. It is worth noting that this relation can be expressed as an equality by introducing an additional error term on the right-hand side, proportional to $\tilde{\eta}_{ij}$.
	Nevertheless, homogeneity imposes a structural assumption on the class of systems under consideration. Such structural assumptions are standard in the data-driven control literature, where many works focus on particular classes of systems, such as polynomial dynamics~\citep{guo2021data}, strict-feedback systems~\citep{basu2026neural}, etc.\hfill $\square$
	\end{remark}}
	In the subsequent definition, we present the concept of \emph{incremental} global asymptotic stability of interconnected dt-NS $\Sigma$.
	
	\begin{definition}\label{GAS}
An interconnected dt-NS $\Sigma$ in~\eqref{Eq:51} is incrementally globally asymptotically stable ($\delta$-GAS) if $$ \Vert x(k) - x'(k)\Vert \leq \beta(\Vert x(0)-x'(0)\Vert ,k),$$
for any $x(0),x'(0)\in \mathbb R^n$ and some $\beta \in \mathcal {KL}$. This indicates that as $k \rightarrow \infty$, any two arbitrary solutions of $\Sigma$ converge to each other towards the origin (which is an equilibrium due to homogeneity).
\end{definition}

We proceed to introduce the next theorem~\citep{angeli}, which elucidates the conditions under which the interconnected dt-NS achieves $\delta$-GAS.
\begin{theorem}\label{GA1S}
Consider an interconnected dt-NS $\Sigma=(\mathbb R^{n},f)$. Let there exist a homogeneous incremental Lyapunov function $\mathcal V:\mathbb R^{n}\times \mathbb R^{n} \rightarrow \mathbb{R}_0^+$ of degree two, {\ie, for any $\eta\in\R^+$ and $x,x'\in\R^n$, $\mathcal V(\eta x,\eta x')=\eta^2\mathcal V(x,x')$, together with constants $\underline{\alpha},\overline\alpha\in\R^+$ and $\gamma\in(0,1)$,} such that
\begin{subequations}
	\begin{align}\notag
		{\forall x,x'\in} &~{\R^n:}\\
		\label{alpha12}
		&\underline{\alpha}\Vert x-x'\Vert^{2} \leq \mathcal V(x,x') \leq \overline{\alpha}\Vert x-x'\Vert^{2},\\\label{alpha1}
		&~~\mathcal V(f(x), f(x')) - \gamma\mathcal V(x,x') \leq 0.
	\end{align}
\end{subequations}
Then the interconnected dt-NS $\Sigma$ exhibits $\delta$-GAS.
\end{theorem}

In the subsequent section, we present $\delta$-ISS properties of individual subsystems.

\section{$\delta$-ISS Properties of Subsystems}\label{Data: ISS}

In this work, the primary aim is to verify the incremental stability of interconnected dt-NS by examining the $\delta$-ISS properties of individual subsystems, formally presented in the subsequent definition.

{\begin{definition}
	A dt-NS $\Sigma_i$ in~\eqref{Eq:2} is incrementally input-to-state stable ($\delta$-ISS) if there exist functions $\beta_i \in \mathcal{KL}$ and $\xi_i \in \mathcal{K}_\infty$ such that
	$$
	\Vert x_i(k) - x'_i(k)\Vert \leq \beta_i(\Vert x_i(0) - x'_i(0)\Vert, k) + \xi_i(\|w_i - w'_i\|_\infty),
	$$
	for all initial conditions $x_i(0), x_i'(0) \in \mathbb{R}^{n_i}$, and all pairs of arbitrary input signals $w_i, w'_i$.
	\end{definition}}
	We now utilize the notion of $\delta$-ISS Lyapunov functions, introduced in the following definition.
	
	\begin{definition}\label{Def:41}
A subsystem $\Sigma_i=(\mathbb R^{n_i},\mathbb R^{p_i},f_i)$ possesses a homogeneous $\delta$-ISS Lyapunov function $\mathcal{S}_i:\mathbb R^{n_i}\times \mathbb R^{n_i}\rightarrow\mathbb{R}_{0}^{+}$  of degree two, if the following constraints are met:
\begin{subequations}
	\begin{align}\notag
		&\forall x_i,x'_i \in \mathbb R^{n_i}\!:\\\label{Eq:8_21}  &~\quad \quad\underline{\alpha}_i\Vert x_i-x'_i\Vert^{{2}} \leq \mathcal S_i(x_i,x'_i) \leq \overline{\alpha}_i\Vert x_i-x'_i\Vert^{{2}},\\\notag
		&\forall x_i,x'_i \in \mathbb R^{n_i},\forall w_i,w'_i \in \mathbb R^{p_i}\!:\\\label{Eq:8_22}
		& \mathcal S_i(f_i(x_i,w_i),f_i(x'_i,w'_i)) \leq \gamma_i\mathcal S_i(x_i,x'_i)+\rho_i\Vert w_i-w'_i\Vert^{{2}}\!,
	\end{align}
\end{subequations}
\!for some $\underline{\alpha}_i,\overline{\alpha}_i\in \mathbb{R}^{+}$, $\rho_i\in\mathbb{R}_{0}^{+}$, and $\gamma_i\in (0,1)$.
\end{definition}

In this work, our primary focus lies in the construction of $\delta$-ISS Lyapunov functions $\mathcal S_i$, aiming to satisfy conditions~\eqref{Eq:8_21}--\eqref{Eq:8_22} locally over the unit sphere $\Vert(x_i,w_i, {x'_i, w'_i})\Vert = 1$. By leveraging the \emph{homogeneity property} of maps $f_i$ and $\delta$-ISS Lyapunov functions $\mathcal S_i$, we then extend these conditions globally to $\mathbb R^{n_i}\times \mathbb R^{p_i}{\times \mathbb R^{n_i}\times \mathbb R^{p_i}}$. Subsequently, in Section~\ref{Guarantee_ROP}, we introduce a compositionality condition derived from small-gain reasoning, which enables the construction of an incremental Lyapunov function $\mathcal V$ for the unknown interconnected network, utilizing $\delta$-ISS Lyapunov functions $\mathcal S_i$ of individual subsystems. Due to the interconnected network's inherent homogeneity, the incremental stability certificate can be guaranteed globally across $\mathbb R^{n}$.

Given the lack of knowledge about the dynamics of each subsystem $f_i$ in our problem, it becomes impossible to search for $\delta$-ISS Lyapunov functions $\mathcal S_i$ that fulfill condition~\eqref{Eq:8_22}. To tackle this key challenge, we introduce a \emph{direct} data-driven approach alongside a compositional condition to construct an incremental Lyapunov function for an unknown interconnected network based on $\delta$-ISS Lyapunov functions of individual subsystems, derived from data. As a result, we formally verify the incremental GAS property of unknown interconnected networks.\vspace{-0.2cm} 

\section{Data-Driven Construction of $\delta$-ISS Lyapunov Functions}\label{Sec:ISS}

In our data-driven methodology, we outline the structure of $\delta$-ISS Lyapunov functions as follows:
\begin{align}\label{Eq:3}
\mathcal{S}_i(q_i,x_i,x'_i)=\sum_{j=1}^{r} {q}_i^j{\varphi_i^j(x_i-x'_i)}, 
\end{align}
where {$\varphi_i^j(x_i-x'_i)$} represents user-defined homogeneous basis functions {of degree two} and $q_i=[{q}_{i}^1;\ldots;q_i^r] \in \mathbb{R}^r$ denotes the unknown variables. It is assumed that $\mathcal{S}_i$ is continuously differentiable.	
{\begin{remark}
	While an immediate choice for basis functions $\varphi_i^j(x_i -x'_i)$ is the quadratic-form family of functions, there are several other families of functions that have non-quadratic forms, such as even-$p$ norm squared functions and (weighted) geometric mean of multiple quadratic forms. Specifically, we define $e_i \coloneq x_i-x'_i$ and describe these basis functions in the following:
	
	(i) Even-$p$ norm squared: One can choose $\varphi_i^j(e_i)=\psi_{i,2m}^j(e_i)$, defined as
	$$
	\psi_{i,2m}^j(e_i)\coloneq\|e_i\|_{2m}^2=\big(\sum_{\ell=1}^{n_i} e_{\ell_i}^{2 m}\big)^{\sfrac{1}{m}}\!,
	$$
	for any $m\geq2$, which results in a non-quadratic $\delta$-ISS Lyapunov function. The degree-two homogeneity is ensured since
	\begin{align*}
		\psi_{i,2m}^j(\eta e_i) &= \|\eta e_i\|_{2m}^2 = \big(\sum_{\ell=1}^{n_i} (\eta e_{\ell_i})^{2m}\big)^{\sfrac{1}{m}}\\
		&= \big(\eta^{2m}\sum_{\ell=1}^{n_i} e_{\ell_i}^{2 m}\big)^{\sfrac{1}{m}}= \eta^2\big(\sum_{\ell=1}^{n_i} e_{\ell_i}^{2 m}\big)^{\sfrac{1}{m}}\\
		&=\eta^2\|e_i\|_{2m}^2=\eta^2\psi_{i,2m}^j(e_i).
	\end{align*}
	
	(ii) (Weighted) geometric mean of multiple quadratic forms: One can opt for $\varphi_i^j(e_i)=\psi_{i,m}^j(e_i)$, described by
	$$
	\psi_{i,m}^j(e_i)\coloneq \big(\prod_{\ell=1}^m e_i^{\top} P_{\ell} e_i\big)^{\sfrac{1}{m}},\; \psi_{i,m}^j(0)=0,
	$$
	for any $m\geq 2$, where $P_\ell\succ 0$ and $P_1\not\propto P_2\not\propto\ldots\not\propto P_\ell$, for all $\ell\in\{1,\ldots,m\}$. The symbol $\not\propto$ implies that two matrices are not proportional to each other, \emph{i.e.}, there exists no $c>0$, such that $P_i=cP_j,\forall i,j\in\{1,\ldots, m\},i\neq j$. This results in a non-quadratic degree-two homogeneous $\delta$-ISS Lyapunov function as
	\begin{align*}
		\psi_{i,m}^j(\eta e_i)&= \big(\prod_{\ell=1}^m (\eta e_i)^{\top} P_{\ell} (\eta e_i)\big)^{\sfrac{1}{m}}=\big(\prod_{\ell=1}^m \eta^2 e_i^{\top} P_{\ell} e_i\big)^{\sfrac{1}{m}}\\
		&=\big(\eta^{2m}\prod_{\ell=1}^m e_i^{\top} P_{\ell} e_i\big)^{\sfrac{1}{m}}=\eta^2 \big(\prod_{\ell=1}^m e_i^{\top} P_{\ell} e_i\big)^{\sfrac{1}{m}}\\
		&=\eta^2	\psi_{i,m}^j(e_i).
	\end{align*}
	This family of functions can even be expressed as a weighted geometric mean of multiple quadratic forms, which is more flexible, as
	$$
	\psi_{i,m}^j(e_i)\coloneq \prod_{\ell=1}^m (e_i^{\top} P_{\ell} e_i)^{a_\ell},\; \psi_{i,m}^j(0)=0,
	$$
	where $a_\ell>0$, with $\sum_{\ell = 1}^{m}a_\ell = 1$. This form is also degree-two homogeneous since
	\begin{align*}
		\psi_{i,m}^j(\eta e_i) &= \prod_{\ell=1}^m ((\eta e_i)^{\top} P_{\ell} (\eta e_i))^{a_\ell}= \prod_{\ell=1}^m (\eta^2 e_i^{\top} P_{\ell} e_i)^{a_\ell}\\
		&=\eta^2 \prod_{\ell=1}^m (e_i^{\top} P_{\ell} e_i)^{a_\ell}=\eta^2	\psi_{i,m}^j(e_i).
	\end{align*}
	It is worth noting that $\psi_{i,m}^j(0)=0$ ensures $\psi_{i,m}^j(e_i)$ is continuously differentiable at zero. \hfill $\square$
	\end{remark}}
	We initiate by formulating the required conditions for the construction of $\delta$-ISS Lyapunov functions, stated in Definition~\ref{Def:41}, as the following robust optimization program (ROP):
\begin{mini!}|s|[2]
{{\substack{\mathcal G_i,\mu_{R_{1_i}}\!,\mu_{R_{2_i}}\\
\mu_{R_{3_i}}\!,\mu_{R_{4_i}}}}}{{\sum_{j = 1}^{4}\mu_{R_{j_i}}},}{\label{ROP}}\notag
\addConstraint{\forall (x_i,w_i,x'_i,w'_i) \in \mathbb R^{n_i}\!\times\! \mathbb R^{p_i}\!\times\! \mathbb R^{n_i}\!\times\! \mathbb R^{p_i}\!\!:}\notag
\addConstraint{\Vert (x_i,w_i,x'_i,w'_i)\Vert = 1},\notag
\addConstraint{\underline{\alpha}_i{\Vert x_i\!-\!x'_i\Vert^2} - \mathcal S_i(q_i,\!x_i,\!x'_i)\notag}
\addConstraint{\hphantom{\underline{\alpha}_i{\Vert x_i-x'_i\Vert^2} -}\leq {\mu_{R_{1_i}}\|x_i - x'_i\|^2}\!,\label{ROP1}}
\addConstraint{\mathcal S_i(q_i,\!x_i,\!x'_i)  -  \overline{\alpha}_i{\Vert x_i\!-\!x'_i\Vert^2}\notag}
\addConstraint{\hphantom{\underline{\alpha}_i{\Vert x_i-x'_i\Vert^2} -}\leq{\mu_{R_{2_i}}\|x_i -x'_i\|^2}\!,\label{ROP2}}
\addConstraint{\mathcal S_i(q_i,\!f_i(x_i,\!w_i),\!f_i(x'_i,\!w'_i))}\notag
\addConstraint{- \gamma_i\mathcal S_i(q_i,\!x_i,\!x'_i)-\rho_i\Vert w_i-w'_i\Vert^2\notag}
\addConstraint{\hphantom{\underline{\alpha}_i{\Vert x_i-x'_i\Vert^2} -}\leq\! {\mu_{R_{3_i}}\|x_i -x'_i\|^2}\!,\label{ROP3}}
\addConstraint{{\rho_i\leq (1-\gamma_i)\mu_{R_{4_i}}},\label{ROP4}}
\addConstraint{\mathcal G_i = [\underline{\alpha}_i;\overline{\alpha}_i;\gamma_i;\rho_i;{q}_{i}^1;\dots;q_{i}^r],\gamma_i\!\in\! (0,1),}\notag
\addConstraint{\underline{\alpha}_i,\!\overline{\alpha}_i \!\in\! [1,\!+\infty),{\mu_{R_{1_i}},\mu_{R_{2_i}},\mu_{R_{3_i}} \!\in\! \mathbb R},}\notag
\addConstraint{\rho_i, {\mu_{R_{4_i}}}\!\in\!\mathbb{R}^{+}_0,q_i^j\in\R, \forall j\in\{1,\dots,r\}.}\notag
\end{mini!}
To ensure the well-posedness of the ROP, we assume that all decision variables in the vector $\mathcal G_i$
are constrained within compact sets{~\citep{esfahani2014performance}. More precisely, since the left-hand side of each constraint depends continuously on $(x_i,w_i,x_i',w_i')$, and they should hold over the compact set $\Vert (x_i,w_i,x'_i,w'_i)\Vert = 1$, for any fixed choice of the decision variables, these quantities remain finite and attain their worst-case values over the unit sphere. In addition, by restricting the decision variables to compact sets, the feasible set is compact, and thus the continuous objective function attains its minimum. Therefore, the problem is well-posed in the sense that an optimizer exists, which is denoted as $\sum_{j = 1}^{4}\mu_{R_{j_i}}^*$ in our work}. When {$\mu^*_{R_{1_i}}\!,\mu^*_{R_{2_i}}\!,\mu^*_{R_{3_i}}$} are {negative}, any feasible solution to the ROP indicates the fulfillment of conditions~\eqref{Eq:8_21}--\eqref{Eq:8_22}, as described in Definition~\ref{Def:41}. {We note that while a unified $\mu_{R_i}$ can be defined for conditions~\eqref{ROP1}--\eqref{ROP3}, as commonly considered in related studies, we instead introduce distinct objectives for these conditions, leading to a potentially less conservative formulation.}
\begin{remark}\label{optimal}
Given that the magnitude of  $\frac{\rho_i}{1-\gamma_i}$ plays a pivotal role in our compositionality condition proposed in Section~\ref{Guarantee_ROP}, we have introduced an additional condition in~\eqref{ROP4}, where the objective involves minimizing {$\mu_{R_{1_i}}$ to $ \mu_{R_{4_i}}$}. {We note, however, that condition~\eqref{ROP4} is not strictly required for the correctness of the proposed framework, and removing it does not invalidate the main theoretical results. Nevertheless, enforcing it during the optimization helps guide the selection of gains toward values that are more favorable for satisfying the small-gain condition~\eqref{Eq:43} in a later stage (cf. Theorem~\ref{Thm:3}), thereby increasing the likelihood that the interconnected system fulfills the compositional condition.}
Note that while the optimal value of {$\sum_{j = 1}^{4}\mu_{R_{j_i}}^*$} is always unique, {$\mu_{R_{1_i}}^*$\!, $\mu_{R_{2_i}}^*$\!, $\mu_{R_{3_i}}^*$\!, and $\mu_{R_{4_i}}^*$} themselves may not be unique. We are particularly interested in the smallest possible {$\mu_{R_{1_i}}^*$\!, $\mu_{R_{2_i}}^*$\!, and $\mu_{R_{3_i}}^*$} among various options that lead to the unique optimal value.\hfill $\square$
\end{remark}

\begin{remark}
Conditions~\eqref{ROP3} and~\eqref{ROP4} in the ROP exhibit bilinearity since $\gamma_i$, $\rho_i$, and $q_i^j$ all are decision variables. To resolve this, we treat $\gamma_i\in(0,1)$ as an element of a finite set with cardinality $l$, denoted as $\gamma_i \in \{\gamma_{i}^1,\dots,\gamma_{i}^l\}$.\hfill $\square$
\end{remark}
The proposed formulation outlined in~\eqref{ROP} is not tractable due to the presence of unknown $f_i$ in condition~\eqref{ROP3} of the ROP. Given this significant hurdle, we offer a data-driven method for constructing $\delta$-ISS Lyapunov functions without the need to directly solve ROP~\eqref{ROP}. To achieve this, {assuming that we can measure all states of the interconnected network (and consequently, those of all subsystems),} we collect sampled data points from trajectories of unknown subsystems, represented as $(\tilde{x}_i^z, \tilde{w}_i^z,\tilde{x}_i^{z'}\!, \tilde{w}_i^{z'}\!, f_i( \tilde{x}_i^z, \tilde{w}_i^z),f_i (\tilde{x}_i^{z'}, \tilde{w}_i^{z'})),$ taken from the set 
$\bar{\mathcal{N}_i}\!: \{ (z,z') \in \{1, \ldots, \mathcal{N}_i \}^2\!\!: z \neq z' \}$, {such that $\tilde x_i^z\neq \tilde x_i^{z'}$ and $\tilde w_i^z\neq \tilde w_i^{z'}$\!.}

In our data-driven setting, we initially normalize all data points by projecting them onto a \emph{unit sphere}. In particular, we collect data points $(\tilde{x}_i^z, \tilde{w}_i^z,\tilde{x}_i^{z'}\!, \tilde{w}_i^{z'}\!, f_i( \tilde{x}_i^z, \tilde{w}_i^z),$ $f_i (\tilde{x}_i^{z'}, \tilde{w}_i^{z'}))$ and normalize each with respect to its initial state and internal input samples using
$\| ( \tilde{x}_i^z, \tilde{w}_i^z,\tilde{x}_i^{z'}, \tilde{w}_i^{z'} ) \|$, \emph{i.e.,}
\begin{align}\notag
	&(
	\hat{x}_i^{z z'},\hat{w}_i^{z z'}, \hat{x}_i^{z'\!z}, \hat{w}_i^{z'\!z}, f_i(\hat{x}_i^{z z'},\hat{w}_i^{z z'}),
	f_i(\hat{x}_i^{z'\!z},\hat{w}_i^{z'\!z})\!) \\\label{New77}
	&=  \qquad \frac{( \tilde{x}_i^z, \tilde{w}_i^z,\tilde{x}_i^{z'}\!, \tilde{w}_i^{z'}\!, f_i( \tilde{x}_i^z, \tilde{w}_i^z),
		f_i (\tilde{x}_i^{z'}, \tilde{w}_i^{z'})\!)}{ \| ( \tilde{x}_i^z, \tilde{w}_i^z,\tilde{x}_i^{z'}, \tilde{w}_i^{z'} ) \| } ,
\end{align}
resulting in the projection onto the unit sphere. {We emphasize that, since $\tilde x_i^z\neq \tilde x_i^{z'}$ and $\tilde w_i^z\neq \tilde w_i^{z'}$ are ensured during sample collection, their projection onto the unit sphere also satisfy $\hat x_i^{zz'}\neq \hat x_i^{z'\!z}$ and $\hat w_i^{zz'}\neq \hat w_i^{z'\!z}$.}
{\begin{remark}
		The proposed framework does not require physically disconnecting the subsystems during data collection. In fact, the data are collected from the interconnected network itself, so that the coupling effects are naturally embedded in the measured trajectories. In particular, the internal input of each subsystem is determined by the states of its neighboring subsystems through the known interconnection structure as described in~\eqref{Eq:41}. Therefore, once state data from the full network are available, $(\hat{x}_i^{z z'},\hat{w}_i^{z z'}, \hat{x}_i^{z'\!z}, \hat{w}_i^{z'\!z})$ required for each subsystem can be directly extracted given the corresponding interconnections.\hfill $\square$
\end{remark}}
Next, we compute the maximum distance between any points on the unit sphere and the dataset as follows:
\begin{align}\notag
	&\varepsilon_i = \max_{\mathcal Q_i}\!\min_{(z,z') \in \bar{\mathcal{N}_i}} \Vert (x_i,\!w_i,\!x'_i,\!w'_i) \!-\! (\hat{x}_i^{z z'}\!\!,\!\hat{w}_i^{z z'}\!\!,\!\hat{x}_i^{z'\!z}\!\!,\!\hat{w}_i^{z'\!z})\Vert,\\\label{EQ:41}
	&{\mathcal Q_i} \in \mathbb R^{n_i}\!\times\! \mathbb R^{p_i}\!\times\! \mathbb R^{n_i}\!\times\! \mathbb R^{p_i}\!\!: \quad\Vert {\mathcal Q_i}\Vert \!=\! 1,
\end{align}
{where $\mathcal Q_i\coloneq (x_i,\!w_i,\!x'_i,\!w'_i)$.} Taking into account $(\hat{x}_i^{z z'},\hat{w}_i^{z z'}, \hat{x}_i^{z'\!z}, \hat{w}_i^{z'\!z})\in \mathbb R^{n_i}\!\times\! \mathbb R^{p_i}\!\times\! \mathbb R^{n_i}\!\times\! \mathbb R^{p_i}$ with  $(z,z') \in \bar{\mathcal{N}_i}$, we put forth the following scenario optimization program (SOP):\vspace{0.1cm}
\begin{mini!}|s|[2]
	{{\substack{\mathcal G_i,\mu_{\mathcal N_{1_i}}\!,\mu_{\mathcal N_{2_i}}\\
	\mu_{\mathcal N_{3_i}}\!,\mu_{\mathcal N_{4_i}}}}}{{\sum_{j = 1}^{4}\mu_{\mathcal N_{j_i}}},}{\label{SOP-1}}\notag
	\addConstraint{\forall (\hat{x}_i^{z z'},\hat{w}_i^{z z'}, \hat{x}_i^{z'\!z}, \hat{w}_i^{z'\!z})\in \mathbb R^{n_i}\!\times\! \mathbb R^{p_i}\!\times\! \mathbb R^{n_i}}\notag
	\addConstraint{\times \mathbb R^{p_i}, \forall (z,z') \in \bar{\mathcal{N}_i},}\notag
	\addConstraint{\underline{\alpha}_i\Vert \hat x^{zz'}_i\!\!-\!\hat x^{z'\!z}_i\Vert^2 \!-\! \mathcal S_i(q_i,\hat x^{zz'}_i\!\!,\hat x^{z'\!z}_i)\notag}
	\addConstraint{\hphantom{\mathcal S_i(q_i,\hat x^{zz'}_i\!\!,\hat x^{z'\!z}_i) \!-\!  \overline{\alpha}_i{\Vert \hat x^{zz'}_i\!\!-}}\leq\! {\mu_{\mathcal N_{1_i}}}\!,\label{SOP1-1}}
	\addConstraint{\mathcal S_i(q_i,\hat x^{zz'}_i\!\!,\hat x^{z'\!z}_i) \!-\!  \overline{\alpha}_i{\Vert \hat x^{zz'}_i\!\!-\!\hat x^{z'\!z}_i\Vert^2}\notag}
	\addConstraint{\hphantom{\mathcal S_i(q_i,\hat x^{zz'}_i\!\!,\hat x^{z'\!z}_i) \!-\!  \overline{\alpha}_i{\Vert \hat x^{zz'}_i\!\!-}}\leq {\mu_{\mathcal N_{2_i}}}\!,\label{SOP2-1}}
	\addConstraint{\mathcal S_i(q_i,\!f_i(\hat x^{zz'}_i\!,\!\hat w^{zz'}_i\!),\!f_i(\hat x^{z'\!z}_i\!,\!\hat w^{z'\!z}_i\!)\!)}\notag
	\addConstraint{ -\! \gamma_i\mathcal S_i(q_i,\!\hat x^{zz'}_i\!\!,\!\hat x^{z'\!z}_i\!)\!-\!\rho_i\Vert \hat w^{zz'}_i\!\!\!-\!\hat w^{z'\!z}_i\Vert^2\notag}
	\addConstraint{\hphantom{\mathcal S_i(q_i,\hat x^{zz'}_i\!\!,\hat x^{z'\!z}_i) \!-\!  \overline{\alpha}_i{\Vert \hat x^{zz'}_i\!\!-}}\leq {\mu_{\mathcal N_{3_i}}}\!,\label{SOP3-1}}
	\addConstraint{{\rho_i\leq (1-\gamma_i)\mu_{\mathcal N_{4_i}}},\label{SOP4-1}}
	\addConstraint{\mathcal G_i = [\underline{\alpha}_i;\overline{\alpha}_i;\gamma_i;\rho_i;{q}_{i}^1;\dots;q_{i}^r], \gamma_i\!\in\! (0,1),}\notag
	\addConstraint{\underline{\alpha}_i,\!\overline{\alpha}_i \!\in\! [1,\!+\infty),{\mu_{\mathcal N_{1_i}},\mu_{\mathcal N_{2_i}},\mu_{\mathcal N_{3_i}} \!\in\! \mathbb R,}}\notag
	\addConstraint{\rho_i, {\mu_{\mathcal N_{4_i}}}\!\in\!\mathbb{R}^{+}_0,q_i^j\in\R, \forall j\in\{1,\dots,r\}.}\notag
\end{mini!}
{It is noteworthy that as we can measure all states of the interconnected network and the interconnection topology is known, the dimensions of both the subsystem states and their internal input signals can be determined directly.} It is evident that the challenge posed by the unknown function $f_i$ in the proposed SOP~\eqref{SOP-1} is effectively addressed since data points $(\tilde{x}_i^z, \tilde{w}_i^z,\tilde{x}_i^{z'}\!, \tilde{w}_i^{z'}\!, f_i( \tilde{x}_i^z, \tilde{w}_i^z),
f_i (\tilde{x}_i^{z'}, \tilde{w}_i^{z'}))$ inherently capture the behavior of the unknown subsystem dynamics. The optimal value of SOP is denoted by {$\sum_{j = 1}^{4}\mu_{\mathcal N_{j_i}}^*$}.
{\begin{remark}
	As the collected samples satisfy $\hat x_i^{zz'}\neq \hat x_i^{z'\!z}$ and $\hat w_i^{zz'}\neq \hat w_i^{z'\!z}$, {it is evident that $\Vert \hat x^{zz'}_i\!-\!\hat x^{z'\!z}_i\Vert\neq 0$, $\Vert \hat w^{zz'}_i-\hat w^{z'\!z}_i\Vert\neq 0$, and $\mathcal S_i(q_i,\cdot,\cdot)\neq 0$ in conditions~\eqref{SOP1-1}--\eqref{SOP3-1}.
	Therefore, the SOP in~\eqref{SOP-1} with $\mu_{\mathcal N_{j_i}}<0$, for $j\in\{1,\dots,3\}$, is not infeasible by construction.}
	Accordingly, the norm terms appearing on the right-hand sides of conditions \eqref{ROP1}–\eqref{ROP3} in the ROP, introduced to ensure well-posedness when $x_i = x_i'$, are no longer required in the corresponding conditions of the SOP~\eqref{SOP-1}. In fact, the SOP and ROP are connected through the left-hand sides of conditions~\eqref{SOP1-1}--\eqref{SOP3-1} and~\eqref{ROP1}--\eqref{ROP3}, which are derived from the original conditions~\eqref{Eq:8_21} and~\eqref{Eq:8_22}. In contrast, the right-hand sides serve as artificial upper bounds, introduced through the optimization objectives to ensure that the original $\delta$-ISS conditions in~\eqref{Eq:8_21} and~\eqref{Eq:8_22} are ultimately satisfied under some newly proposed conditions (see~\eqref{Con3-1}–\eqref{Con3-3} and the proof of Theorem~\ref{Thm2}).$\hfill\square$
\end{remark}}
{Although the SOP in~\eqref{SOP-1} can be solved using data, it requires the exact measurements of $f_i(\hat x^{zz'}_i\!,\!\hat w^{zz'}_i\!)$ and $\!f_i(\hat x^{z'\!z}_i\!,\!\hat w^{z'\!z}_i\!)$, which are not accessible. More precisely, as illustrated in Fig.~\ref{fig:noise}, we assume that the one-step transition data are corrupted by $\varsigma_i$, defined as
	\begin{subequations}\label{eq:noisy measurements}
		\begin{align}
			\hat f_i(\hat x^{zz'}_i\!,\!\hat w^{zz'}_i\!) &= f_i(\hat x^{zz'}_i\!,\!\hat w^{zz'}_i\!) + \varsigma_i,\\
			\hat f_i(\hat x^{z'\!z}_i\!,\!\hat w^{z'\!z}_i\!) &= f_i(\hat x^{z'\!z}_i\!,\!\hat w^{z'\!z}_i\!) + \varsigma'_i,
		\end{align}
	\end{subequations}
	where $\varsigma'_i$ is a different realization of the same disturbance. Hence, the available data for the one-step transitions are $\hat f_i(\hat x^{zz'}_i\!,\!\hat w^{zz'}_i\!)$ and $\hat f_i(\hat x^{z'\!z}_i\!,\!\hat w^{z'\!z}_i\!)$. We consider $\varsigma_i$ as measurement noise under the following assumption.
	\begin{figure}[t!]
		\centering
		\includegraphics[width=0.8\linewidth]{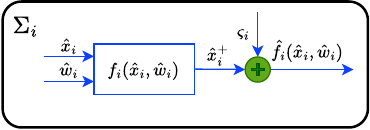}
		\caption{{The one-step ahead state is corrupted by an unknown bounded noise $\varsigma_i$.}}
		\label{fig:noise}
	\end{figure}
	\begin{assumption}\label{asmp:noise}
		The measurement noise $\varsigma_i$ is unknown, but its norm is bounded by some known bounds $\mathsf d$, satisfying $\|\varsigma_i\|\leq \mathsf d$.
\end{assumption}}
{We note that this is a standard assumption in the literature on data-driven methods with noisy measurements, as it enables robust analysis while capturing practical sensing uncertainties. In order to solve the problem using the noise-corrupted data, we aim to propose a noise-influenced SOP, denoted by $\text{SOP}_\varsigma$, which only relies on the accessible noisy measurements and thereby can be solved. Then, we aim to show that the solution of $\text{SOP}_\varsigma$ is always a valid solution for the original SOP~\eqref{SOP-1}. To do so, we first propose the following lemma, showing the linear growth of gradient of the candidate $\delta$-ISS Lyapunov function with respect to its arguments.}
{\begin{lemma}\label{lem:growth}
		Consider a degree-two homogeneous $\delta$-ISS Lyapunov function $\bar{\mathcal S}_i(e_i) \coloneq \mathcal S_i(q_i,x_i,x_i')$, with $e_i \coloneq x_i-x_i'$. Then, for any fixed $q_i \in \R^{r}$, there always exists a constant $L_i>0$ such that
		\begin{subequations}\label{eq:grad and L}
			\begin{align}\label{eq:grad bound}
				\Big\|\nabla \bar{\mathcal S}_i(e_i)\Big\|
				\leq L_i\|e_i\|, \qquad \forall e_i\in\R^{n_i},
			\end{align}
			where
			\begin{align}\label{eq:Lq}
				L_i\coloneq \sup_{\|e_i\|=1}\|\nabla\bar{\mathcal S}_i(e_i)\|.
			\end{align}
		\end{subequations}
\end{lemma}}
{\textbf{Proof.} To demonstrate that~\eqref{eq:grad bound} holds, we first show that $\nabla \bar{\mathcal S}_i(e_i)$ is homogeneous of degree one. According to~\eqref{Eq:3}, since $\bar{\mathcal S}_i(e_i)$ is homogeneous of degree two, we have
	\begin{align}\label{eq:hom S}
		\bar{\mathcal S}_i(\eta e_i) = \eta^2\bar{\mathcal S}_i(e_i), \quad \forall \eta>0, \forall e_i\in\R^{n_i}.
	\end{align}
	Now, by taking derivative from the left-hand side of~\eqref{eq:hom S} with respect to $e_i$ and using the chain rule, we obtain
	$$
	\nabla\bar{\mathcal S}_i(\eta e_i) = \eta \nabla\bar{\mathcal S}_i(\eta e_i).
	$$
	Similarly, taking the derivative of the right-hand side of~\eqref{eq:hom S} with respect to $e_i$ results in
	$$
	\nabla(\eta^2\bar{\mathcal S}_i(e_i)) =  \eta^2\nabla\bar{\mathcal S}_i(e_i).
	$$
	Then, through the equality in~\eqref{eq:hom S}, one can readily see that
	$$
	\eta \nabla\bar{\mathcal S}_i(\eta e_i) = \eta^2\nabla\bar{\mathcal S}_i(e_i)\Rightarrow \nabla\bar{\mathcal S}_i(\eta e_i) = \eta\nabla\bar{\mathcal S}_i(e_i),
	$$
	which implies the degree-one homogeneity of $\nabla\bar{\mathcal S}_i(e_i)$.
	Now, let us define
	$
	L_i\coloneq \sup_{\|e_i\|=1}\|\nabla\bar{\mathcal S}_i(e_i)\|.
	$
	Since $\bar{\mathcal S}_i(e_i)$ is continuously differentiable, $\nabla\bar{\mathcal S}_i(e_i)$ is always continuous. Hence, $\|\nabla\bar{\mathcal S}_i(e_i)\|$ is continuous on the compact unit sphere $\{e_i\in\R^{n_i}\!\!:\!\|e_i\| \!=\! 1\}$, and consequently, $L_i$ is finite.
	
	Then, by defining $e_i = \|e_i\|\frac{e_i}{\|e_i\|}$ for $e_i\neq 0$, and using the homogeneity property of $\nabla\bar{\mathcal S}_i(e_i)$, we obtain
	$$
	\nabla\bar{\mathcal S}_i(e_i)=\nabla\bar{\mathcal S}_i\big(\|e_i\|\tfrac{e_i}{\|e_i\|}\big)=\|e_i\|\, \nabla \bar{\mathcal S}_i\big(\tfrac{e_i}{\|e_i\|}\big).
	$$
	Subsequently, by taking norms from both sides, we attain
	$$
	\|\nabla\bar{\mathcal S}_i(e_i)\| = \|e_i\|\, \big\|\nabla \bar{\mathcal S}_i\big(\tfrac{e_i}{\|e_i\|}\big)\big\|\overset{\eqref{eq:Lq}}{\leq}L_i\|e_i\|,
	$$
	which also holds immediately for $e_i=0$, and thereby completes the proof.$\hfill\blacksquare$}

{Lemma~\ref{lem:growth} ensures that, for a degree-two homogeneous $\mathcal S_i(q_i,x_i,x'_i)$, there always exists $L_i$ satisfying~\eqref{eq:grad bound}. In order to construct the $\delta$-ISS Lyapunov function $\mathcal S_i(q_i,x_i,x'_i)$ in our setting, the constant $L_i$ is required~(cf. constraint~\eqref{SOP3-1-n}). However, the value of $L_i$ itself depends on $\mathcal S_i(q_i,x_i,x'_i)$, which is not designed yet. Hence, we aim to propose an approach to construct both $L_i$ and  $\mathcal S_i(q_i,x_i,x'_i)$, simultaneously.}

{To do so, we first recall the definition of $\mathcal S_i(q_i,x_i,x'_i)$ in~\eqref{Eq:3} in terms of $e_i=x_i-x'_i$ as
	\begin{align}\label{eq:S e}
		\bar{\mathcal S}_i(e_i)=\sum_{j=1}^{r} {q}_i^j\varphi_i^j(e_i).
	\end{align}
	Taking the derivative of~\eqref{eq:S e}, and subsequently taking norms, results in $\|\nabla\bar{\mathcal S}_i(e_i)\|=\big\|\sum_{j=1}^{r} {q}_i^j\nabla \varphi_i^j(e_i)\big\|$. Then, according to~\eqref{eq:Lq}, we have
	\begin{align*}
		L_i=\sup_{\|e_i\|=1}\|\nabla\bar{\mathcal S}_i(e_i)\|=\sup_{\|e_i\|=1}\big\|\sum_{j=1}^{r} {q}_i^j\nabla \varphi_i^j(e_i)\big\|.
	\end{align*}
	Using the triangle inequality and the Cauchy-Schwarz inequality~\citep{bhatia1995cauchy}, one gets
	\begin{align}\label{newoj}
		L_i\leq\sum_{j=1}^{r} \|{q}_i^j\|\sup_{\|e_i\|=1}\|\nabla \varphi_i^j(e_i)\|.
	\end{align}
	We define $\bar L_i^j\coloneq\sup_{\|e_i\|=1}\|\nabla \varphi_i^j(e_i)\|$ and introduce auxiliary variables {$\bar q_i^j\geq0$}, such that
	\begin{align}\label{eq:aux q}
		\bar q_i^j \geq q_i^j,\quad \bar q_i^j \geq -q_i^j,
	\end{align}
	which ensures $\bar q_i^j\geq \|q_i^j\|$. Considering~\eqref{newoj}, it yields
	\begin{align*}
		L_i\leq\sum_{j=1}^{r} \bar q_i^j\bar L_i^j.
	\end{align*}
	Therefore, imposing
	\begin{align}\label{eq:L hat}
		\hat L_i\geq\sum_{j=1}^{r} \bar q_i^j\bar L_i^j,
	\end{align}
	provides an upper bound on $L_i$ and satisfies~\eqref{eq:grad bound}. It is worth noting that $\bar L_i^j$ can be computed a priori as the basis functions $\varphi_i^j(e_i)$ are chosen before solving $\text{SOP}_{\varsigma}$. 
	
	We provide the following illustrative example, elucidating the computation of $\bar L_i^j$.}

{\begin{example}\label{example}
		Take $n_i = 2$ and let $\varphi_i^1(e_i) = e_{1_i}^2$, $\varphi_i^2(e_i) = e_{1_i}e_{2_i}$, and $\varphi_i^3(e_i) = e_{2_i}^2$. Then,
		$$
		\nabla \varphi_i^1(e_i)=\begin{bmatrix}
			2e_{1_i}\\
			0
		\end{bmatrix}\!, \nabla \varphi_i^2(e_i)=\begin{bmatrix}
			e_{2_i}\\
			e_{1_i}
		\end{bmatrix}\!, \nabla \varphi_i^3(e_i)=\begin{bmatrix}
			0\\
			2e_{2_i}
		\end{bmatrix}\!.
		$$
		To obtain $\bar L_i^1$, one can take the norm of $\nabla \varphi_i^1(e_i)$ and obtain
		$\|\nabla \varphi_i^1(e_i)\| = 2\|e_{1_i}\|$. Now, we restrict our attention to $\{e_i:\|e_i\| = 1\}$, which is equivalent to $e_{1_i}^2 + e_{2_i}^2 = 1$. Then, $\|\nabla \varphi_i^1(e_i)\|$ attains its maximum at $e_i = [\pm 1;0]$, resulting in $\|\nabla \varphi_i^1(e_i)\| = 2\|e_{1_i}\|\leq 2$. Hence, $\bar L_i^1= 2$. Similarly, one can readily obtain $\bar L_i^3 = 2$. To compute $\bar L_i^2$, we take norm of $\nabla \varphi_i^2(e_i)$, which yields
		$$
		\|\nabla \varphi_i^2(e_i)\| = \sqrt{e_{1_i}^2 + e_{2_i}^2} = \|e_i\|.
		$$
		Consequently, on the unit sphere $\{e_i:\|e_i\| = 1\}$, we have $\|\nabla \varphi_i^2(e_i)\| = 1$, implying that $\bar L_i^2 = 1$.
	\end{example}
	Having established Lemma~\ref{lem:growth}, we propose $\text{SOP}_{\varsigma}$, incorporating additional conditions~\eqref{eq:aux q} and \eqref{eq:L hat} as constraints, as follows.}
{\begin{mini!}|s|[2]
		{\substack{\bar{\mathcal G}_i,\mu_{\varsigma_{1_i}}\!,\mu_{\varsigma_{2_i}}\\
		\mu_{\varsigma_{3_i}}\!,\mu_{\varsigma_{4_i}}}}{\sum_{j = 1}^{4}\mu_{\varsigma_{j_i}} ,}{\label{SOP-1-n}}\notag
		\addConstraint{\forall (\hat{x}_i^{z z'},\hat{w}_i^{z z'}, \hat{x}_i^{z'\!z}, \hat{w}_i^{z'\!z})\in \mathbb R^{n_i}\!\times\! \mathbb R^{p_i}\!\times\! \mathbb R^{n_i}}\notag
		\addConstraint{\times \mathbb R^{p_i}, \forall (z,z') \!\in\! \bar{\mathcal{N}_i},}\notag
		\addConstraint{\underline{\alpha}_i\Vert \hat x^{zz'}_i\!\!-\!\hat x^{z'\!z}_i\Vert^2 \!-\! \mathcal S_i(q_i,\hat x^{zz'}_i\!\!,\hat x^{z'\!z}_i)\notag}
		\addConstraint{\hphantom{\mathcal S_i(q_i,\hat x^{zz'}_i\!\!,\hat x^{z'\!z}_i) \!-\!  \overline{\alpha}_i{\Vert \hat x^{zz'}_i\!\!-}}\leq\! \mu_{\varsigma_{1_i}}\!,\label{SOP1-1-n}}
		\addConstraint{\mathcal S_i(q_i,\hat x^{zz'}_i\!\!,\hat x^{z'\!z}_i) \!-\!  \overline{\alpha}_i{\Vert \hat x^{zz'}_i\!\!-\!\hat x^{z'\!z}_i\Vert^2}\notag}
		\addConstraint{\hphantom{\mathcal S_i(q_i,\hat x^{zz'}_i\!\!,\hat x^{z'\!z}_i) \!-\!  \overline{\alpha}_i{\Vert \hat x^{zz'}_i\!\!-}}\leq\! \mu_{\varsigma_{2_i}}\!,\label{SOP2-1-n}}
		\addConstraint{\mathcal S_i(q_i,\!\hat f_i(\hat x^{zz'}_i\!,\!\hat w^{zz'}_i\!),\!\hat f_i(\hat x^{z'\!z}_i\!,\!\hat w^{z'\!z}_i\!)\!)}\notag
		\addConstraint{ -\! \gamma_i\mathcal S_i(q_i,\!\hat x^{zz'}_i\!\!,\!\hat x^{z'\!z}_i\!)\!-\!\rho_i\Vert \hat w^{zz'}_i\!\!\!-\!\hat w^{z'\!z}_i\Vert^2\notag}
		\addConstraint{+\!2\mathsf d \hat L_i\big(\|\hat f_i(\hat x^{zz'}_i\!,\!\hat w^{zz'}_i\!)-\!\hat f_i(\hat x^{z'\!z}_i\!,\!\hat w^{z'\!z}_i\!)\|\!+\!2\mathsf d\big)}\notag
		\addConstraint{\hphantom{\mathcal S_i(q_i,\hat x^{zz'}_i\!\!,\hat x^{z'\!z}_i) \!-\!  \overline{\alpha}_i{\Vert \hat x^{zz'}_i\!\!-}}\leq\! \mu_{\varsigma_{3_i}}\!,\label{SOP3-1-n}}
		\addConstraint{q_i^j - \bar q_i^j\leq 0,\quad -q_i^j - \bar q_i^j\leq 0,}\label{SOP5-1-n}
		\addConstraint{-\hat L_i + \sum_{j=1}^{r} \bar q_i^j\bar L_i^j\leq 0,}\label{SOP6-1-n}
		\addConstraint{\rho_i\leq (1-\gamma_i)\mu_{\varsigma_{4_i}},\label{SOP4-1-n}}
		\addConstraint{\bar{\mathcal G}_i \!=\! [\underline{\alpha}_i;\overline{\alpha}_i;\gamma_i;\rho_i;{q}_{i}^1;\dots;q_{i}^r;\bar q_i^1;\dots;\bar q_i^r;\!\hat L_i],}\notag
		\addConstraint{\gamma_i\!\in\! (0,1),\underline{\alpha}_i,\!\overline{\alpha}_i \!\in\! \![1,\!+\infty), \!\mu_{\varsigma_{1_i}},\mu_{\varsigma_{2_i}},\mu_{\varsigma_{3_i}} \!\in\! \mathbb R,}\notag
		\addConstraint{\rho_i,\! \mu_{\varsigma_{4_i}}\!\!\in\!\mathbb{R}^{+}_0\!,q_i^j\in\R,\bar{q}_i^j\!\in\!\R_0^+\!, \forall j\!\in\!\{1,\dots,r\},}\notag
		\addConstraint{\hat L_i\in\R^+\!.}\notag
	\end{mini!}
	
	As can be seen, $\text{SOP}_\varsigma$ in~\eqref{SOP-1-n} relies only on the corrupted data, which now makes it tractable. It is evident that when $\mathsf d = 0$, \emph{i.e.}, $\hat f_i(\cdot,\cdot) = f_i(\cdot,\cdot)$ in~\eqref{eq:noisy measurements} without any measurement noise, condition~\eqref{SOP3-1-n} in $\mathrm{SOP}_{\varsigma}$~\eqref{SOP-1-n} recovers condition~\eqref{SOP3-1} in SOP~\eqref{SOP-1}. We denote the optimal value of  $\text{SOP}_\varsigma$ by {$\sum_{j = 1}^{4}\mu_{\varsigma_{j_i}}^*$}\!.}

{\begin{remark}
		While the homogeneity property potentially restricts the class of nonlinear systems under consideration, it enables us to derive \emph{global} guarantees over the entire Euclidean space. More specifically, we normalize the collected data and project them onto the unit sphere. We then solve $\text{SOP}_\varsigma$ in~\eqref{SOP-1-n} (whose solution is a valid solution for SOP~\eqref{SOP-1}, as demonstrated in Lemma~\ref{lem:sol}) over the compact set defined by $\|(\hat{x}_i^{z z'},\hat{w}_i^{z z'}, \hat{x}_i^{z'\!z}, \hat{w}_i^{z'\!z})\| = 1,$ and establish correctness via conditions~\eqref{Con3-1}--\eqref{Con3-3}, which ensure that the solution of $\text{SOP}_\varsigma$ guarantees the solution of the ROP in~\eqref{ROP}. By exploiting the homogeneity property, we subsequently lift the obtained results from the unit sphere to the entire state space. While this assumption could be relaxed, this would come at the expense of losing global guarantees and restricting the analysis to \emph{local} results over specific compact subsets.\hfill $\square$
\end{remark}}
{We introduce the following lemma, showing that the solution of $\text{SOP}_\varsigma$~\eqref{SOP-1-n} is also a valid solution for SOP~\eqref{SOP-1}.}

{\begin{lemma}\label{lem:sol}
	Consider individual subsystems $\Sigma_i = (\R^{n_i},\R^{p_i},f_i)$ in~\eqref{Eq:2}, each with a candidate $\delta$-ISS Lyapunov function $\mathcal S_i(q_i,x_i,x'_i)$ as in~\eqref{Eq:3}. Let Assumption~\ref{asmp:noise} hold. Under Lemma~\ref{lem:growth}, any solution of $\text{SOP}_{\varsigma}$ in~\eqref{SOP-1-n} is a valid solution for SOP~\eqref{SOP-1}.
\end{lemma}}
{\textbf{Proof.} For any $(z,z')\in\bar{\mathcal N}_i$, we define
	\begin{align}
		&\begin{array}{l}
			\varpi_i^{zz'}\coloneq f_i(\hat{x}_i^{z z'},\hat{w}_i^{z z'}) - f_i(\hat{x}_i^{z'\!z},\hat{w}_i^{z'\!z}),\\
			\hat{\varpi}_i^{zz'}\coloneq \hat f_i(\hat{x}_i^{z z'},\hat{w}_i^{z z'}) - \hat f_i(\hat{x}_i^{z'\!z},\hat{w}_i^{z'\!z}),
		\end{array}\label{eq:varpi}\\
		&\begin{array}{l}
			\bar{\mathcal S}_i(\varpi_i^{zz'})\coloneq \mathcal S_i(q_i,f_i(\hat{x}_i^{z z'},\hat{w}_i^{z z'}),f_i(\hat{x}_i^{z'\!z},\hat{w}_i^{z'\!z})),\\
			\bar{\mathcal S}_i(\hat\varpi_i^{zz'})\coloneq \mathcal S_i(q_i,\hat f_i(\hat{x}_i^{z z'},\hat{w}_i^{z z'}),\hat f_i(\hat{x}_i^{z'\!z},\hat{w}_i^{z'\!z})).
		\end{array}\label{eq:SbarVarpi}
	\end{align}
	Now, consider the line segment joining $\varpi_i^{zz'}$ and $\hat{\varpi}_i^{zz'}$ as
	\begin{align}\label{eq:segment}
		\chi_i(h_i)\!\coloneq \hat{\varpi}_i^{zz'}\!\!+h_i(\varpi_i^{zz'}\!\!-\!\hat{\varpi}_i^{zz'}), \quad h_i\in[0,1],
	\end{align}
	indicating that as $h_i$ moves from zero to one, $\chi_i(h_i)$ moves from $\hat{\varpi}_i^{zz'}$ to $\varpi_i^{zz'}$, \ie, $\chi_i(0)=\hat\varpi_i^{zz'}$ and $\chi_i(1)=\varpi_i^{zz'}$. We take the norm of $\chi_i(h_i)$, and by applying the triangle inequality, obtain
	\begin{align*}
		\|\chi_i(h_i)\|&= \|\hat{\varpi}_i^{zz'}+h_i(\varpi_i^{zz'}\!-\hat{\varpi}_i^{zz'})\|\\
		&\leq \|\hat{\varpi}_i^{zz'}\|+h_i\|\varpi_i^{zz'}\!-\hat{\varpi}_i^{zz'}\|.
	\end{align*}
	Since $h_i\in[0, 1]$, one can deduce that
	\begin{align}\label{eq:norm chi}
		\|\chi_i(h_i)\|\leq \|\hat{\varpi}_i^{zz'}\|+\|\varpi_i^{zz'}\!-\hat{\varpi}_i^{zz'}\|.
	\end{align}
	Furthermore, according to Assumption~\ref{asmp:noise}, using the triangle inequality, one can obtain
	\begin{align}
		\|\varpi_i^{zz'}\!-\hat{\varpi}_i^{zz'}\|&\overset{\eqref{eq:varpi}}{=}\|(f_i(\hat{x}_i^{z z'},\hat{w}_i^{z z'}) -\hat f_i(\hat{x}_i^{z z'},\hat{w}_i^{z z'})\!) \notag\\
		&\hphantom{\overset{\eqref{eq:noisy measurements}}{=} \|}- ( f_i(\hat{x}_i^{z'\!z},\hat{w}_i^{z'\!z}) - \hat f_i(\hat{x}_i^{z'\!z},\hat{w}_i^{z'\!z})\!)\|\notag\\
		&\overset{\hphantom{\eqref{eq:noisy measurements}}}{\leq}\|f_i(\hat{x}_i^{z z'},\hat{w}_i^{z z'}) -\hat f_i(\hat{x}_i^{z z'},\hat{w}_i^{z z'})\|\notag\\
		&\hphantom{\overset{\eqref{eq:noisy measurements}}{=}} + \| f_i(\hat{x}_i^{z'\!z},\hat{w}_i^{z'\!z}) - \hat f_i(\hat{x}_i^{z'\!z},\hat{w}_i^{z'\!z})\|\notag\\
		&\overset{\eqref{eq:noisy measurements}}{=} \|\varsigma_i\| +\|\varsigma'_i\|\leq2\mathsf d.\label{eq:bound varpi}
	\end{align}
	Substituting~\eqref{eq:bound varpi} into~\eqref{eq:norm chi} yields
	\begin{align}\label{eq:bound chi}
		\|\chi_i(h_i)\|\leq \|\hat{\varpi}_i^{zz'}\|+2\mathsf d.
	\end{align}
	Using the Fundamental Theorem of Calculus followed by the chain rule, one has
	\begin{align}
		\bar{\mathcal S}_i(\chi_i(1))&-\bar{\mathcal S}_i(\chi_i(0))=\bar{\mathcal S}_i(\varpi_i^{zz'})-\bar{\mathcal S}_i(\hat{\varpi}_i^{zz'})\notag \\
		&=\!\!\int_0^1 \!\frac{\mathrm d}{\mathrm d h_i}\bar{\mathcal S}_i(\chi_i(h_i))\mathrm d h_i \notag\\
		&= \!\!\int_0^1\!\nabla \bar{\mathcal S}_i(\chi_i(h_i))^{\!\top}\frac{\partial \chi_i(h_i)}{\partial h_i}\mathrm d h_i\notag\\
		&\overset{\eqref{eq:segment}}{=} \!\!\int_0^1\!\nabla \bar{\mathcal S}_i(\chi_i(h_i))^{\!\top} \!(\varpi_i^{zz'}\!\!-\!\hat{\varpi}_i^{zz'}\!)\mathrm d h_i.\label{eq:FTC-1}
	\end{align}
	Now, by taking the norm of both sides of this expression and using the Cauchy-Schwarz inequality, one gets
	\begin{align}
		\|\bar{\mathcal S}_i(\varpi_i^{zz'})&-\bar{\mathcal S}_i(\hat{\varpi}_i^{zz'})\|\notag\\
		&\overset{\hphantom{\eqref{eq:hom S}}}{\leq}\! \int_0^1\!\!\|\nabla \bar{\mathcal S}_i(\chi_i(h_i))^{\!\top} \!(\varpi_i^{zz'}\!\!\!-\!\hat{\varpi}_i^{zz'}\!)\|\mathrm d h_i\notag\\
		&\overset{\hphantom{\eqref{eq:hom S}}}{\leq}\! \int_0^1\!\!\|\nabla \bar{\mathcal S}_i(\chi_i(h_i))\|\|\varpi_i^{zz'}\!\!\!-\!\hat{\varpi}_i^{zz'}\|\mathrm d h_i\notag\\
		&\overset{\hphantom{\eqref{eq:hom S}}}{=}\|\varpi_i^{zz'}\!\!\!-\!\hat{\varpi}_i^{zz'}\| \int_0^1\!\!\|\nabla \bar{\mathcal S}_i(\chi_i(h_i))\|\mathrm d h_i\label{eq:FTC-2}
	\end{align}
	Then, under Lemma~\ref{lem:growth}, we have
	\begin{align}
		\|\bar{\mathcal S}_i(\varpi_i^{zz'})&-\bar{\mathcal S}_i(\hat{\varpi}_i^{zz'})\|\notag\\
		&\overset{\eqref{eq:grad and L}}{\leq} \|\varpi_i^{zz'}\!\!\!-\!\hat{\varpi}_i^{zz'}\| \int_0^1\!\!L_i\|\chi_i(h_i)\|\mathrm d h_i\notag\\
		&\!\overset{\eqref{eq:bound varpi}}{\leq} 2\mathsf dL_i \int_0^1\!\!\|\chi_i(h_i)\|\mathrm d h_i\notag\\
		&\!\overset{\eqref{eq:bound chi}}{\leq} 2\mathsf dL_i \int_0^1\!\! (\|\hat{\varpi}_i^{zz'}\|+2\mathsf d) \mathrm d h_i\notag\\
		&\overset{\hphantom{\eqref{eq:hom S}}}{=}2\mathsf dL_i (\|\hat{\varpi}_i^{zz'}\|+2\mathsf d),\label{eq:FTC-3}
	\end{align}
	implying that
	\begin{align*}
		\bar{\mathcal S}_i(\varpi_i^{zz'})&\overset{\hphantom{\eqref{eq:L hat}}}{\leq}\bar{\mathcal S}_i(\hat{\varpi}_i^{zz'}) + 2\mathsf dL_i (\|\hat{\varpi}_i^{zz'}\|+2\mathsf d)\\
		&\overset{\eqref{eq:L hat}}{\leq}\bar{\mathcal S}_i(\hat{\varpi}_i^{zz'}) + 2\mathsf d\hat L_i (\|\hat{\varpi}_i^{zz'}\|+2\mathsf d).
	\end{align*}
	Through substitution of the definitions of $\hat{\varpi}_i^{zz'}$, $\bar{\mathcal S}_i(\varpi_i^{zz'})$, and $\bar{\mathcal S}_i(\hat{\varpi}_i^{zz'})$ into this inequality, and by adding $- \gamma_i\mathcal S_i(q_i,\!\hat x^{zz'}_i\!\!,\!\hat x^{z'\!z}_i\!)\!-\!\rho_i\Vert \hat w^{zz'}_i\!\!\!-\!\hat w^{z'\!z}_i\Vert^2$
	to both sides, we obtain
	\begin{align}
		&\mathcal S_i(q_i,\!f_i(\hat{x}_i^{z z'}\!\!,\!\hat{w}_i^{z z'}\!),f_i(\hat{x}_i^{z'\!z}\!,\!\hat{w}_i^{z'\!z})\!)- \gamma_i\mathcal S_i(q_i,\!\hat x^{zz'}_i\!\!,\!\hat x^{z'\!z}_i\!)\notag\\
		&-\!\rho_i\Vert \hat w^{zz'}_i\!\!\!-\!\hat w^{z'\!z}_i\Vert^2\notag\\
		&\leq\mathcal S_i(q_i,\!\hat f_i(\hat{x}_i^{z z'}\!\!,\!\hat{w}_i^{z z'}\!),\hat f_i(\hat{x}_i^{z'\!z}\!,\!\hat{w}_i^{z'\!z})\!) - \gamma_i\mathcal S_i(q_i,\!\hat x^{zz'}_i\!\!,\!\hat x^{z'\!z}_i\!)\notag\\
		&-\!\rho_i\Vert \hat w^{zz'}_i\!\!\!-\!\hat w^{z'\!z}_i\Vert^2\notag\\
		&+ 2\mathsf d\hat L_i (\|\hat f_i(\hat{x}_i^{z z'}\!\!,\!\hat{w}_i^{z z'}\!) \!-\! \hat f_i(\hat{x}_i^{z'\!z}\!,\!\hat{w}_i^{z'\!z})\|+2\mathsf d).\label{eq:upper}
	\end{align}
	Now, if one enforces
	\begin{align*}
		&\mathcal S_i(q_i,\!\hat f_i(\hat{x}_i^{z z'}\!\!,\!\hat{w}_i^{z z'}\!),\hat f_i(\hat{x}_i^{z'\!z}\!,\!\hat{w}_i^{z'\!z})\!) - \gamma_i\mathcal S_i(q_i,\!\hat x^{zz'}_i\!\!,\!\hat x^{z'\!z}_i\!)\\
		&-\!\rho_i\Vert \hat w^{zz'}_i\!\!\!-\!\hat w^{z'\!z}_i\Vert^2\\
		&+ 2\mathsf d\hat L_i (\|\hat f_i(\hat{x}_i^{z z'}\!\!,\!\hat{w}_i^{z z'}\!) \!-\! \hat f_i(\hat{x}_i^{z'\!z}\!,\!\hat{w}_i^{z'\!z})\|+2\mathsf d)\leq\mu_{\varsigma_{3_i}},
	\end{align*}
	which is the proposed constraint in~\eqref{SOP3-1-n}, as per~\eqref{eq:upper}, constraint~\eqref{SOP3-1} in SOP~\eqref{SOP-1} becomes upper bounded by $\mu_{\varsigma_{3_i}}$. Therefore, choosing $\mu_{\mathcal N_{3_i}}\coloneq \mu_{\varsigma_{3_i}}^*$ is valid and ensures the satisfaction of constraint~\eqref{SOP3-1} with solutions $\bar{\mathcal G}_i^*$ from $\text{SOP}_{\varsigma}$.
	
	It is noteworthy that the left-hand sides of constraints~\eqref{SOP1-1-n} and \eqref{SOP2-1-n} are identical to those of constraints~\eqref{SOP1-1} and \eqref{SOP2-1}. As the solution of $\text{SOP}_{\varsigma}$ fulfills constraints~\eqref{SOP1-1-n} and \eqref{SOP2-1-n} {with $\mu_{\varsigma_{1_i}}^*$ and $\mu_{\varsigma_{2_i}}^*$, respectively, one can readily deduce that choosing $\mu_{\mathcal N_{1_i}}\coloneq\mu_{\varsigma_{1_i}}^*$ and $\mu_{\mathcal N_{2_i}}\coloneq\mu_{\varsigma_{2_i}}^*$ also ensures constraints~\eqref{SOP1-1} and \eqref{SOP2-1} hold, respectively}, with solutions $\bar{\mathcal G}_i^*$ from $\text{SOP}_{\varsigma}$, which concludes the proof.$\hfill\blacksquare$}

In the next section, we employ the proposed {$\text{SOP}_\varsigma$} and construct $\delta$-ISS Lyapunov functions for unknown $\Sigma_i$  with a guaranteed correctness.

\section{Correctness Guarantees}
{In this work, we assume that each unknown subsystem $f_i(x_i,w_i)$ is Lipschitz continuous concerning $(x_i,w_i)$. Consequently, the unknown noisy measurement map $\hat f_i(\cdot,\cdot)$ is Lipschitz continuous as well, with Lipschitz constant $\mathscr L_i^{\hat f}$ (cf. Remark~\ref{New1}). Since our analysis is confined to the unit sphere (bounded domain), given the continuously differentiable nature of $\delta$-ISS Lyapunov functions, $\mathcal S_i^*(q_i,\cdot,\cdot) = \mathcal S_i(q_i^*,\cdot,\cdot)$ is also Lipschitz continuous with respect to $(x_i,x'_i)$, with a Lipschitz constant denoted by $\mathscr{L}^{\mathcal S}_{i}$. Similarly, it follows that $\underline{\alpha}_i^*{\Vert x_i-x'_i\Vert^2} - \mathcal S_i^*(q_i,x_i,x'_i)$ and $\mathcal S_i^*(q_i,x_i,x'_i) - \overline{\alpha}_i^*{\Vert x_i-x'_i\Vert^2}$ are both Lipschitz continuous with respect to $(x_i,x'_i)$, with Lipschitz constants $\mathscr{L}^1_{i}$ and $\mathscr{L}^2_{i}$, respectively. At a later stage in Subsection \ref{Lipschitz}, we propose an algorithm to estimate the Lipschitz constant $\mathscr L_i^{\hat f}$ for unknown subsystems based on data, while we compute $\mathscr{L}^{\mathcal S}_{i}, \mathscr{L}^1_{i}$, and $\mathscr{L}^2_{i}$ analytically. {We emphasize that $\hat f_i(\cdot,\cdot)$ remains bounded since its arguments are on the unit sphere, which is a bounded domain. Therefore, its global Lipschitz constant always exists.}}

{Utilizing Lemma~\ref{lem:sol}, which shows that the solution of $\text{SOP}_\varsigma$~\eqref{SOP-1-n} is also a valid solution for SOP~\eqref{SOP-1}, we now present one of the main contributions of our work in the following theorem, ensuring the satisfaction of the ROP~\eqref{ROP} by providing correctness guarantees.}
{\begin{theorem}\label{Thm2}
	Consider individual subsystems in~\eqref{Eq:2} and their corresponding noise-corrupted $\text{SOP}_\varsigma$ in~\eqref{SOP-1-n} with associated optimal values $\mu_{\varsigma_{1_i}}^*$, $\mu_{\varsigma_{2_i}}^*$, $\mu_{\varsigma_{3_i}}^*$, and solutions $\bar{\mathcal G}_i^* \!\!=\!\! [\underline{\alpha}_i^*;\overline{\alpha}_i^*;\gamma_i^*;\rho_i^*;{q}^{1*}_{i};\dots;$ $q^{r*}_{i};\bar q_i^{1*};\dots;\bar q_i^{r*};\!\hat L_i^*]$ with $\bar{\mathcal N}_i$. Under Lemma~\ref{lem:sol}, if, for all $i\in\{1,\dots,\mathcal M\}$,
	\begin{subequations}\label{eq:rob cons}
		\begin{align}
				&\mu^*_{\varsigma_{1_i}}+ \mathscr L_i^1\varepsilon_i  \leq 0,\label{Con3-1}\\
				&\mu^*_{\varsigma_{2_i}}+ \mathscr L_i^2\varepsilon_i  \leq 0,\label{Con3-2}\\
				&\mu^*_{\varsigma_{3_i}} + \mathscr L^3_i\varepsilon_i  \leq 0,\label{Con3-3}
		\end{align}
	\end{subequations}
	where
	\begin{align}
		\mathscr L^3_i=&~4\mathscr L_i^{\hat f}L_i(2\mathsf d + \mathscr L_i^{\hat f}\varepsilon_i)+ \gamma_i^*\mathscr L_i^{\mathcal S} + {4\sqrt{2}}\rho_i^*\notag\\
		&+2\mathscr L_i^{\hat f}L_i\max_{\bar{\mathcal N}_i}\|\hat f_i(\hat{x}_i^{zz'}\!\!,\!\hat{w}_i^{zz'}\!) \!-\! \hat f_i(\hat{x}_i^{z'\!z}\!,\!\hat{w}_i^{z'\!z})\|,\label{eq:L3}
	\end{align}
	\!\!then the constructed $\mathcal{S}_i^*$ obtained by solving $\text{SOP}_\varsigma$~\eqref{SOP-1-n} serve as $\delta$-ISS Lyapunov functions for unknown subsystems $\Sigma_i$ with correctness guarantees.
\end{theorem}}

{\bf Proof.} We demonstrate that under the conditions in~\eqref{eq:rob cons}, one can conclude the satisfaction of the ROP in~\eqref{ROP} while providing correctness guarantees.

Initially, we demonstrate that under condition~\eqref{Con3-1}, the functions $\mathcal{S}_i^*$ obtained by solving {$\text{SOP}_\varsigma$ in~\eqref{SOP-1-n}} satisfy the lower bound of condition~\eqref{Eq:8_21} across the entire state space, with {$\mu_{\varsigma_{1_i}}^*< 0$}, \emph{i.e.,}
\begin{align*}\notag
	\underline{\alpha}_i^*\Vert x_i-x'_i\Vert^2 - \mathcal S_i^*(q_i,x_i,x'_i) \leq 0.
\end{align*}
For given $(x_i,x_i')$, let $(\bar z, \bar{z}' ) \!\coloneq\!\argmin_{(z,z') \in \bar{\mathcal{N}_i}} \Vert (x_i,x'_i) - (\hat{x}_i^{z z'}\!,\hat{x}_i^{z'\!z})\Vert$. By incorporating the term $\underline{\alpha}_i^*{\Vert \hat x^{\bar z\bar z'}_i-\hat x^{\bar z'\!\bar z}_i\Vert^2} - \mathcal S_i^*(q_i,\hat x^{\bar z\bar z'}_i,\hat x^{\bar z'\!\bar z}_i)$ through both addition and subtraction, we have
\begin{align*}
	\underline{\alpha}_i^*&\Vert x_i-x'_i\Vert^2\!\! -\!\mathcal S_i^*\!(q_i,x_i,x'_i) = \underline{\alpha}_i^*\Vert x_i-x'_i\Vert^2 \!\!-\!\mathcal S_i^*\!(q_i,x_i,x'_i)\\
	& -(\underline{\alpha}_i^*{\Vert  \hat x^{\bar z\bar z'}_i-\hat x^{\bar z'\!\bar z}_i\Vert^2} - \mathcal S_i^*(q_i,\hat x^{\bar z\bar z'}_i,\hat x^{\bar z'\!\bar z}_i))\\
	& +~~\! \underline{\alpha}_i^*{\Vert  \hat x^{\bar z\bar z'}_i-\hat x^{\bar z'\!\bar z}_i\Vert^2} - \mathcal S_i^*(q_i,\hat x^{\bar z\bar z'}_i,\hat x^{\bar z'\!\bar z}_i).
\end{align*}
{We define $\mathcal Q_i\coloneq(x_i,\!w_i,\!x'_i,\!w'_i)$ and use this notation only where the space is limited.} Given that $\underline{\alpha}_i^*{\Vert x_i-x'_i\Vert^2} - \mathcal S_i^*(q_i,x_i,x'_i)$ is Lipschitz continuous with respect to $(x_i,x'_i)$ with a Lipschitz constant $\mathscr{L}^1_{i}$, and since
\begin{align*}
	&\underset{(x_i,x'_i): ~\!\|(x_i,x'_i)\| \leq 1}\max \| (x_i,x'_i) - (\hat{x}_i^{z z'}\!,\hat{x}_i^{z'\!z}) \| \\
	&=\underset{\mathcal Q_i: \Vert \mathcal Q_i\Vert = 1}\max \| (x_i,x'_i) - (\hat{x}_i^{z z'}\!,\hat{x}_i^{z'\!z})  \|, \text{and} \\
	&\| (x_i,\!x'_i) \!-\! (\hat{x}_i^{z z'}\!\!,\!\hat{x}_i^{z'\!z})  \| \leq \Vert (x_i,\!w_i,\!x'_i,\!w'_i) \!-\! (\hat{x}_i^{z z'}\!\!,\!\hat{w}_i^{z z'}\!\!,\!\hat{x}_i^{z'\!z}\!\!,\!\hat{w}_i^{z'\!z})\Vert,
\end{align*}
we have
\begin{align*}
	&\underline{\alpha}_i^*\Vert x_i-x'_i\Vert^2 -\mathcal S_i^*(q_i,x_i,x'_i)\\  
	&\leq\mathscr{L}^1_{i} \, \min_{\bar{\mathcal{N}_i}}\Vert (x_i,x'_i) - (\hat{x}_i^{z z'}\!,\hat{x}_i^{z'\!z}) \Vert\\
	&~~~+ \underline{\alpha}_i^*{\Vert \hat x^{\bar z\bar z'}_i-\hat x^{\bar z'\!\bar z}_i\Vert^2} - \mathcal S_i^*(q_i,\hat x^{\bar z\bar z'}_i,\hat x^{\bar z'\!\bar z}_i)\\
	&\leq \mathscr{L}^1_{i}\! \max_{(x_i,x'_i)}\min_{\bar{\mathcal{N}_i}}\Vert (x_i,x'_i) - (\hat{x}_i^{z z'}\!,\hat{x}_i^{z'\!z}) \Vert\\
	&~~~+ \underline{\alpha}_i^*{\Vert  \hat x^{\bar z\bar z'}_i-\hat x^{\bar z'\!\bar z}_i\Vert^2} - \mathcal S_i^*(q_i,\hat x^{\bar z\bar z'}_i,\hat x^{\bar z'\!\bar z}_i)\\
	&= \mathscr{L}^1_{i}\max_{\mathcal Q_i}\min_{\bar{\mathcal{N}_i}}\Vert (x_i,x'_i) - (\hat{x}_i^{z z'}\!,\hat{x}_i^{z'\!z}) \Vert\\
	&~~~+ \underline{\alpha}_i^*{\Vert  \hat x^{\bar z\bar z'}_i-\hat x^{\bar z'\!\bar z}_i\Vert^2} - \mathcal S_i^*(q_i,\hat x^{\bar z\bar z'}_i,\hat x^{\bar z'\!\bar z}_i)\\
	&\leq \mathscr{L}^1_{i}\max_{\mathcal Q_i}\min_{\bar{\mathcal{N}_i}}\Vert(x_i,\!w_i,\!x'_i,\!w'_i) \!-\! (\hat{x}_i^{z z'}\!\!,\!\hat{w}_i^{z z'}\!\!,\!\hat{x}_i^{z'\!z}\!\!,\!\hat{w}_i^{z'\!z}) \Vert\\
	&~~~+ \underline{\alpha}_i^*{\Vert  \hat x^{\bar z\bar z'}_i-\hat x^{\bar z'\!\bar z}_i\Vert^2} - \mathcal S_i^*(q_i,\hat x^{\bar z\bar z'}_i,\hat x^{\bar z'\!\bar z}_i)\\
	&\!\!\!\overset{\eqref{EQ:41}}{=} \!\!\mathscr{L}_{i}^1 \varepsilon_i \!+\! \underline{\alpha}_i^*{\Vert  \hat x^{\bar z\bar z'}_i-\hat x^{\bar z'\!\bar z}_i\Vert^2} \!-\! \mathcal S_i^*(q_i,\hat x^{\bar z\bar z'}_i,\hat x^{\bar z'\!\bar z}_i)
\end{align*}

According to {condition~\eqref{SOP1-1-n} of $\text{SOP}_\varsigma$}, we have
$$
\underline{\alpha}_i^*\Vert x_i-x'_i\Vert^2 -\mathcal S_i^*(q_i,x_i,x'_i)\leq {\mu_{\varsigma_{1_i}}^* \!+\! \mathscr{L}_{i}^1 \varepsilon_{i}}.
$$
Given our proposed condition in~\eqref{Con3-1}, one can conclude that
\begin{align*}
	\underline{\alpha}_i^*\Vert x_i-x'_i\Vert^2 -\mathcal S_i^*(q_i,x_i,x'_i)\leq 0.
\end{align*}
By leveraging a similar rationale, it can be shown that under condition~\eqref{Con3-2}, the functions $\mathcal{S}_i^*$ obtained by solving {$\text{SOP}_\varsigma$ in~\eqref{SOP-1-n}} satisfy the upper bound of condition~\eqref{Eq:8_21} across the entire state space with {$\mu_{\varsigma_{2_i}}^*<0$}.

We proceed with showing that the constructed $\mathcal{S}_i^*$ obtained by solving {$\text{SOP}_\varsigma$ in~\eqref{SOP-1-n}} satisfies condition~\eqref{Eq:8_22} across the state and internal input spaces, with {$\mu_{\varsigma_{3_i}}^*< 0$}, \emph{i.e.,}
\begin{align}
	&\mathcal S_i^*(q_i,f_i(x_i,w_i),f_i(x'_i,w'_i)) -\gamma_i^*\mathcal S_i^*(q_i,x_i,x'_i)\notag\\
	&-\rho_i^*\Vert w_i\!-\!w'_i\Vert^2\!\leq\! 0,\notag\\
	&\forall {\mathcal Q_i }\in \mathbb R^{n_i}\times \mathbb R^{p_i}\times \mathbb R^{n_i}\times \mathbb R^{p_i}\!\!:\quad
	\Vert {\mathcal Q_i}\Vert = 1.\label{eq:unit sphere}
\end{align}
For given $(x_i,w_i,x_i',w_i')$ and with a slight abuse of notation, let us define
\begin{align}\label{eq:barzz'}
	(\bar z, \bar{z}' ) \!\coloneq\!\argmin_{\bar{\mathcal{N}_i}} \Vert (x_i,\!w_i,\!x'_i,\!w'_i) \!-\! (\hat{x}_i^{z z'}\!\!,\!\hat{w}_i^{z z'}\!\!,\!\hat{x}_i^{z'\!z}\!\!,\!\hat{w}_i^{z'\!z})\Vert,
\end{align}
{where $(z,z') \in \bar{\mathcal{N}_i}$}. According to Remark~\ref{optimal}, our interest lies in the pair $(\bar z, \bar{z}')$ that yields the smallest possible {$\mu_{\varsigma_{3_i}}^*$}. By incorporating the terms {$\mathcal S_i^*(q_i,\!\hat f_i(\hat x^{\bar z\bar z'}_i\!,\!\hat w^{\bar z\bar z'}_i\!),$ $\!\hat f_i(\hat x^{\bar z'\!\bar z}_i,\!\hat w^{\bar z'\!\bar z}_i)\!)- \gamma_i^*\mathcal S_i^*(q_i,\hat x^{\bar z\bar z'}_i\!,\hat x^{\bar z'\!\bar z}_i)\!-\!\rho_i^*\Vert \hat w^{\bar z\bar z'}_i\!-\!\hat w^{\bar z'\!\bar z}_i\Vert^2$} through addition and subtraction, one has
\begin{align}
	&\mathcal S_i^*(q_i,f_i(x_i,w_i),f_i(x'_i,w'_i)\!) \!-\!\gamma_i^*\mathcal S_i^*(q_i,x_i,x'_i)\!-\!\rho_i^*\Vert w_i\!-\!w'_i\Vert^2\notag\\
	& =\!\mathcal S_i^*(q_i,\!f_i(x_i,\!w_i),\!f_i(x'_i,\!w'_i)\!) \!-\!\gamma_i^*\mathcal S_i^*(q_i,\!x_i,\!x'_i)\!-\!\rho_i^*\Vert w_i\!-\!w'_i\Vert^2\notag\\
	&~~~{-(\mathcal S_i^*(q_i,\!\hat f_i(\hat x^{\bar z\bar z'}_i\!\!,\!\hat w^{\bar z\bar z'}_i),\!\hat f_i(\hat x^{\bar z'\!\bar z}_i\!\!,\!\hat w^{\bar z'\!\bar z}_i)\!) - \gamma_i^*\mathcal S_i^*(q_i,\hat x^{\bar z\bar z'}_i\!\!,\hat x^{\bar z'\!\bar z}_i)}\notag\\
	&~~~{-\rho_i^*\Vert  \hat w^{\bar z\bar z'}_i-\hat w^{\bar z'\!\bar z}_i\Vert^2)}\notag\\
	&~~~{+\mathcal S_i^*(q_i,\!\hat f_i(\hat x^{\bar z\bar z'}_i\!\!,\!\hat w^{\bar z\bar z'}_i),\!\hat f_i(\hat x^{\bar z'\!\bar z}_i\!\!,\!\hat w^{\bar z'\!\bar z}_i)\!) - \gamma_i^*\mathcal S_i^*(q_i,\hat x^{\bar z\bar z'}_i\!\!,\hat x^{\bar z'\!\bar z}_i)}\notag\\
	&~~~{-\rho_i^*\Vert  \hat w^{\bar z\bar z'}_i-\hat w^{\bar z'\!\bar z}_i\Vert^2.}\label{eq:con 3 SOP}
\end{align}
{We begin by deriving an upper bound on $\mathcal S_i^*(q_i,f_i(x_i,w_i),$ $f_i(x'_i,w'_i)\!)-\mathcal S_i^*(q_i,\!\hat f_i(\hat x^{\bar z\bar z'}_i\!\!,\!\hat w^{\bar z\bar z'}_i),\!\hat f_i(\hat x^{\bar z'\!\bar z}_i\!\!,\!\hat w^{\bar z'\!\bar z}_i)\!)$. To do so, we first obtain a bound on $\|(f_i(x_i,w_i) -f_i(x'_i,w'_i))- (\hat f_i(\hat x^{\bar z\bar z'}_i\!\!,\!\hat w^{\bar z\bar z'}_i) - \hat f_i(\hat x^{\bar z'\!\bar z}_i\!\!,\!\hat w^{\bar z'\!\bar z}_i))\|$. Through addition and subtraction of term $\hat f_i(x_i,w_i) -\hat f_i(x'_i,w'_i)$ inside the norm, and using triangle inequality afterwards, we obtain
	\begin{align*}
		&\|(f_i(x_i,\!w_i) \!-\!f_i(x'_i,\!w'_i)\!)\!-\! (\hat f_i(\hat x^{\bar z\bar z'}_i\!\!,\!\hat w^{\bar z\bar z'}_i) \!-\! \hat f_i(\hat x^{\bar z'\!\bar z}_i\!\!,\!\hat w^{\bar z'\!\bar z}_i)\!)\|\\
		&=\!\|f_i(x_i,\!w_i)\!-\! \hat f_i(x_i,\!w_i)\!-\!(f_i(x'_i,\!w'_i) \!-\! \hat f_i(x'_i,\!w'_i)\!) \\
		&\hphantom{=}+ \!\hat f_i(x_i,\!w_i) \!-\! \hat f_i(\hat x^{\bar z\bar z'}_i\!\!,\!\hat w^{\bar z\bar z'}_i) \!-\! (\hat f_i(x'_i,\!w'_i) \!-\! \hat f_i(\hat x^{\bar z'\!\bar z}_i\!\!,\!\hat w^{\bar z'\!\bar z}_i)\!) \|\\
		&\leq\underbrace{\|f_i(x_i,\!w_i)\!-\! \hat f_i(x_i,\!w_i)\|}_{a_1}\!+\!\underbrace{\|f_i(x'_i,\!w'_i) \!-\! \hat f_i(x'_i,\!w'_i)\|}_{a_2} \\
		&\hphantom{\leq}\!+ \underbrace{\|\hat f_i(x_i,\!w_i) \!-\! \hat f_i(\hat x^{\bar z\bar z'}_i\!\!,\!\hat w^{\bar z\bar z'}_i)\|}_{a_3} \!+\! \underbrace{\|\hat f_i(x'_i,\!w'_i) \!- \!\hat f_i(\hat x^{\bar z'\!\bar z}_i\!\!,\!\hat w^{\bar z'\!\bar z}_i)\|}_{a_4}.
	\end{align*}
	One can obtain an upper bound on $a_1$ and $a_2$ via Assumption~\ref{asmp:noise} as
	\begin{align*}
		&\|(f_i(x_i,\!w_i) \!-\!f_i(x'_i,\!w'_i)\!)\!-\! (\hat f_i(\hat x^{\bar z\bar z'}_i\!\!,\!\hat w^{\bar z\bar z'}_i) \!-\! \hat f_i(\hat x^{\bar z'\!\bar z}_i\!\!,\!\hat w^{\bar z'\!\bar z}_i)\!)\|\\
		&\overset{\eqref{eq:noisy measurements}}{\leq}\mathsf d+\mathsf d + a_3 +a_4=2\mathsf d + a_3 + a_4.
	\end{align*}
	Owing to the fact that map $\hat f_i(\cdot,\cdot)$ is Lipschitz continuous with Lipschitz constant $\mathscr L_i^{\hat f}$, as its arguments lie within a bounded domain and $f_i(\cdot,\cdot)$ is Lipschitz continuous, we can obtain upper bounds concerning terms $a_3$ and $a_4$ by utilizing the definition of Lipschitz continuity. Hence, we have
	\begin{align*}
		&\|(f_i(x_i,\!w_i) \!-\!f_i(x'_i,\!w'_i)\!)\!-\! (\hat f_i(\hat x^{\bar z\bar z'}_i\!\!,\!\hat w^{\bar z\bar z'}_i) \!-\! \hat f_i(\hat x^{\bar z'\!\bar z}_i\!\!,\!\hat w^{\bar z'\!\bar z}_i)\!)\|\\
		&\leq \!2\mathsf d \!+\! \mathscr L_i^{\hat f}\!\|\!(x_i,\!w_i)\!-\!(\hat x^{\bar z \bar z'}_i\!\!,\!\hat w^{\bar z \bar z'}_i\!)\!\|\!+\!\mathscr L_i^{\hat f}\!\|\!(x'_i,\!w'_i)\!-\!(\hat x^{\bar z'\!\bar z}_i\!\!,\!\hat w^{\bar z'\!\bar z}_i)\!\|.
	\end{align*}
	Since $\|(x_i,\!w_i)-(\hat x^{\bar z \bar z'}_i\!\!,\!\hat w^{ \bar z \bar z'}_i\!)\|\leq \|(x_i,\!w_i,\!x'_i,\!w'_i) -(\hat{x}_i^{\bar z \bar z'}\!\!,\!\hat{w}_i^{\bar z \bar z'}\!\!,$ $\!\hat{x}_i^{\bar z'\!\bar z}\!\!,\!\hat{w}_i^{\bar z'\!\bar z})\|$ and $\|(x'_i,\!w'_i)-(\hat x^{\bar z'\!\bar z}_i\!\!,\!\hat w^{ \bar z'\!\bar z}_i\!)\|\leq \|(x_i,\!w_i,\!x'_i,\!w'_i) -(\hat{x}_i^{\bar z \bar z'}\!\!,\!\hat{w}_i^{\bar z \bar z'}\!\!,\!\hat{x}_i^{\bar z'\!\bar z}\!\!,\!\hat{w}_i^{\bar z'\!\bar z})\|$, using the definition of $(\bar z,\bar z')$ in~\eqref{eq:barzz'}, one obtains
	\begin{align*}
		&\|(f_i(x_i,\!w_i) \!-\!f_i(x'_i,\!w'_i)\!)\!-\! (\hat f_i(\hat x^{\bar z\bar z'}_i\!\!,\!\hat w^{\bar z\bar z'}_i) \!-\! \hat f_i(\hat x^{\bar z'\!\bar z}_i\!\!,\!\hat w^{\bar z'\!\bar z}_i)\!)\|\\
		&\!\!\overset{\eqref{eq:barzz'}}{\leq} \!\!2\mathsf d \!+ \!2\mathscr L_i^{\hat f}\!\min_{ \bar{\mathcal{N}_i}}\!\|(x_i,\!w_i,\!x'_i,\!w'_i) \!-\! (\hat{x}_i^{z z'}\!\!,\!\hat{w}_i^{z z'}\!\!,\!\hat{x}_i^{z'\!z}\!\!,\!\hat{w}_i^{z'\!z}\!)\|\\
		&\!\overset{\hphantom{\eqref{EQ:41}}}{\leq}\!\!2\mathsf d \!+\! 2\mathscr L_i^{\hat f}\!\max_{\mathcal Q_i}\min_{\bar{\mathcal{N}_i}}\!\|\!(x_i,\!w_i,\!x'_i,\!w'_i) \!-\! (\hat{x}_i^{z z'}\!\!,\!\hat{w}_i^{z z'}\!\!,\!\hat{x}_i^{z'\!z}\!\!,\!\hat{w}_i^{z'\!z}\!)\!\|\\
		&\!\overset{\eqref{EQ:41}}{=}\!\!2\mathsf d \!+\! 2\mathscr L_i^{\hat f}\varepsilon_i.
	\end{align*}
	Now that a bound on $\|(f_i(x_i,w_i) -f_i(x'_i,w'_i))- (\hat f_i(\hat x^{\bar z\bar z'}_i\!\!,\!\hat w^{\bar z\bar z'}_i) - \hat f_i(\hat x^{\bar z'\!\bar z}_i\!\!,\!\hat w^{\bar z'\!\bar z}_i))\|$ is obtained, we can adopt a similar approach as in the proof of Lemma~\ref{lem:sol} and derive a bound on $\mathcal S_i^*(q_i,f_i(x_i,w_i),f_i(x'_i,w'_i)\!)-\mathcal S_i^*(q_i,\!\hat f_i(\hat x^{\bar z\bar z'}_i\!\!,\!\hat w^{\bar z\bar z'}_i),\!\hat f_i(\hat x^{\bar z'\!\bar z}_i\!\!,\!\hat w^{\bar z'\!\bar z}_i)\!)$. More precisely, we define a line segment from $\hat{\varpi}_i^{\bar z\bar z'}\coloneq \hat f_i(\hat x^{\bar z\bar z'}_i\!\!,\!\hat w^{\bar z\bar z'}_i) - \hat f_i(\hat x^{\bar z'\!\bar z}_i\!\!,\!\hat w^{\bar z'\!\bar z}_i)$ to $\varpi_i\coloneq f_i(x_i,w_i) -f_i(x'_i,w'_i)$, and obtain the upper bound on its norm as $\|\hat{\varpi}_i^{\bar z \bar z'}\| + 2\mathsf d \!+\! 2\mathscr L_i^{\hat f}\varepsilon_i$, following~\eqref{eq:segment}--\eqref{eq:bound chi}. Then, utilizing the Fundamental Theorem of Calculus over the line segment and under Lemma~\ref{lem:growth}, following~\eqref{eq:FTC-1}--\eqref{eq:FTC-3}, we obtain
	\begin{align}
		&\mathcal S_i^*\!(q_i,\! f_i(x_i,\!w_i),\!f_i(x'_i,\!w'_i)\!) \! - \! \mathcal S_i^*\!(q_i,\!f_i(\hat x^{\bar z\bar z'}_i\!\!,\!\hat w^{\bar z\bar z'}_i\!),\!f_i(\hat x^{\bar z'\!\bar z}_i\!\!,\!\hat w^{\bar z'\!\bar z}_i)\!)\notag\\
		&\leq(2\mathsf d + 2\mathscr L_i^{\hat f}\varepsilon_i)L_i\big(\|\hat f_i(\hat{x}_i^{\bar z \bar z'}\!\!,\!\hat{w}_i^{\bar z \bar z'}\!) \!-\! \hat f_i(\hat{x}_i^{\bar z'\!\bar z}\!,\!\hat{w}_i^{\bar z'\!\bar z})\| \notag\\
		&\hphantom{\leq(}+ 2\mathsf d + 2\mathscr L_i^{\hat f}\varepsilon_i\big)\notag\\
		&\leq4\mathscr L_i^{\hat f}\varepsilon_iL_i(2\mathsf d + \mathscr L_i^{\hat f}\varepsilon_i ) \notag\\
		&\hphantom{\leq}+2\mathscr L_i^{\hat f}\varepsilon_iL_i\|\hat f_i(\hat{x}_i^{\bar z \bar z'}\!\!,\!\hat{w}_i^{\bar z \bar z'}\!) \!-\! \hat f_i(\hat{x}_i^{\bar z'\!\bar z}\!,\!\hat{w}_i^{\bar z'\!\bar z})\|\notag\\
		&\hphantom{\leq}+2\mathsf d L_i\big(\|\hat f_i(\hat{x}_i^{\bar z \bar z'}\!\!,\!\hat{w}_i^{\bar z \bar z'}\!) \!-\! \hat f_i(\hat{x}_i^{\bar z'\!\bar z}\!,\!\hat{w}_i^{\bar z'\!\bar z})\| + 2\mathsf d\big) .\label{eq:b1}
	\end{align}
	It is worth noting that, since the solution $\bar{\mathcal G}_i$ is obtained via solving the $\text{SOP}_{\varsigma}$~\eqref{SOP-1-n}, and thereby, $q_i^*$ is determined, one can compute $L_i$ analytically according to~\eqref{eq:Lq}.}

{We now continue with providing an upper bound on $\gamma_i^*(\mathcal S_i^*(q_i,\hat x^{\bar z\bar z'}_i\!\!,\hat x^{\bar z'\!\bar z}_i)-\mathcal S_i^*(q_i,x_i,x'_i))$ as follows:
	\begin{align}
		&\gamma_i^*(\mathcal S_i^*(q_i,\hat x^{\bar z\bar z'}_i\!\!,\hat x^{\bar z'\!\bar z}_i)-\mathcal S_i^*(q_i,x_i,x'_i))\notag\\
		&\overset{\hphantom{\eqref{eq:barzz'}}}{\leq}\!\!\gamma_i^*\mathscr L_i^{\mathcal S}\|(x_i,x'_i)-(\hat x^{\bar z \bar z'}_i\!\!,\hat x^{\bar z'\! \bar z}_i)\|\notag\\
		&\overset{\eqref{eq:barzz'}}{\leq}\!\!\gamma_i^*\mathscr L_i^{\mathcal S}\min_{\bar{\mathcal N}_i}\|(x_i,\!w_i,\!x'_i,\!w'_i) \!-\! (\hat{x}_i^{z z'}\!\!,\!\hat{w}_i^{z z'}\!\!,\!\hat{x}_i^{z'\!z}\!\!,\!\hat{w}_i^{z'\!z})\|\notag\\
		&\overset{\hphantom{\eqref{eq:barzz'}}}{\leq}\!\!\gamma_i^*\mathscr L_i^{\mathcal S}\!\max_{\mathcal Q_i}\min_{\bar{\mathcal N}_i}\|\!(x_i,\!w_i,\!x'_i,\!w'_i) \!-\! (\hat{x}_i^{z z'}\!\!,\!\hat{w}_i^{z z'}\!\!,\!\hat{x}_i^{z'\!z}\!\!,\!\hat{w}_i^{z'\!z})\!\|\notag\\
		&\overset{\eqref{EQ:41}}{=}\gamma_i^*\mathscr L_i^{\mathcal S}\varepsilon_i.\label{eq:b2}
\end{align}}
{Finally, we should derive an upper bound on $\rho_i^*(\|w_i\!-\!w'_i\|^2 - \|\hat w^{\bar z\bar z'}_i-\hat w^{\bar z'\!\bar z}_i\|^2)$. We know
	\begin{align*}
		&\rho_i^*(\|w_i\!-\!w'_i\|^2 - \|\hat w^{\bar z\bar z'}_i-\hat w^{\bar z'\!\bar z}_i\|^2)\\
		&\!\!=\!\rho_i^*\underbrace{\big(\|w_i\!-\!w'_i\| \!+\! \|\hat w^{\bar z\bar z'}_i\!\!-\!\hat w^{\bar z'\!\bar z}_i\|\big)}_{(\spadesuit)}\big(\|w_i\!-\!w'_i\| \!-\! \|\hat w^{\bar z\bar z'}_i\!\!-\!\hat w^{\bar z'\!\bar z}_i\|\big).
	\end{align*}
	{Since, as per~\eqref{eq:unit sphere}, the analysis is carried out over the unit sphere and all data pairs are normalized, \ie, $\|(x_i,w_i,x'_i,w'_i)\|=1$ and $\|(\hat{x}_i^{z z'}\!\!,\!\hat{w}_i^{z z'}\!\!,\!\hat{x}_i^{z'\!z}\!\!,\!\hat{w}_i^{z'\!z})\|=1$, it follows that $\|w_i - w'_i\|\leq \sqrt{2}$ and $\|\hat w^{\bar z\bar z'}_i -\hat w^{\bar z'\!\bar z}_i\|\leq \sqrt{2}$.  Hence, $(\spadesuit)$ can be upper bounded as}
	\begin{align*}
		\rho_i^*(\|w_i\!-\!w'_i\|^2 &- \|\hat w^{\bar z\bar z'}_i-\hat w^{\bar z'\!\bar z}_i\|^2)\\
		&\leq{2\sqrt{2}}\rho_i^*\underbrace{\big(\|w_i\!-\!w'_i\| \!-\! \|\hat w^{\bar z\bar z'}_i\!\!-\!\hat w^{\bar z'\!\bar z}_i\|\big)}_{(\blacklozenge)}.
	\end{align*}
	We also use the triangle inequality on $(\blacklozenge)$ twice, yielding
	\begin{align}
		&\rho_i^*(\|w_i\!-\!w'_i\|^2 - \|\hat w^{\bar z\bar z'}_i-\hat w^{\bar z'\!\bar z}_i\|^2)\notag\\
		&\overset{\hphantom{\eqref{eq:barzz'}}}{\leq}\!\!{2\sqrt{2}}\rho_i^*\big(\|w_i\!-\!\hat w^{\bar z\bar z'}_i \!+\! \hat w^{\bar z'\!\bar z}_i \!- \! w'_i\|\big)\notag\\
		&\overset{\hphantom{\eqref{eq:barzz'}}}{\leq}\!\!{2\sqrt{2}}\rho_i^*\big(\|w_i\!-\!\hat w^{\bar z\bar z'}_i \|\!+\!\| w'_i\!-\!\hat w^{\bar z'\!\bar z}_i\|\big)\notag\\
		&\overset{\eqref{eq:barzz'}}{\leq}\!\!{4\sqrt{2}}\rho_i^*\min_{\bar{\mathcal N}_i}\|(x_i,\!w_i,\!x'_i,\!w'_i) \!-\! (\hat{x}_i^{z z'}\!\!,\!\hat{w}_i^{z z'}\!\!,\!\hat{x}_i^{z'\!z}\!\!,\!\hat{w}_i^{z'\!z})\|\notag\\
		&\overset{\hphantom{\eqref{eq:barzz'}}}{\leq}\!\!{4\sqrt{2}}\rho_i^*\max_{\mathcal Q_i}\min_{\bar{\mathcal N}_i}\|(x_i,\!w_i,\!x'_i,\!w'_i) \!-\! (\hat{x}_i^{z z'}\!\!,\!\hat{w}_i^{z z'}\!\!,\!\hat{x}_i^{z'\!z}\!\!,\!\hat{w}_i^{z'\!z})\|\notag\\
		&\overset{\eqref{EQ:41}}{=}{4\sqrt{2}}\rho_i^*\varepsilon_i.\label{eq:b3}
\end{align}
Now, we substitute upper bounds \eqref{eq:b1}--\eqref{eq:b3} into~\eqref{eq:con 3 SOP}, which yields}
	{\begin{align*}
		&\mathcal S_i^*(q_i,f_i(x_i,w_i),f_i(x'_i,w'_i)\!) \!-\!\gamma_i^*\mathcal S_i^*(q_i,x_i,x'_i)\!-\!\rho_i^*\Vert w_i\!-\!w'_i\Vert^2\\
		& \leq4\mathscr L_i^{\hat f}\varepsilon_iL_i(2\mathsf d + \mathscr L_i^{\hat f}\varepsilon_i)+ \gamma_i^*\mathscr L_i^{\mathcal S}\varepsilon_i + {4\sqrt{2}}\rho_i^*\varepsilon_i\\
		&~~+2\mathscr L_i^{\hat f}\varepsilon_iL_i\|\hat f_i(\hat{x}_i^{\bar z \bar z'}\!\!,\!\hat{w}_i^{\bar z \bar z'}\!) \!-\! \hat f_i(\hat{x}_i^{\bar z'\!\bar z}\!,\!\hat{w}_i^{\bar z'\!\bar z})\| \\
		&~~+2\mathsf d L_i\big(\|\hat f_i(\hat{x}_i^{\bar z \bar z'}\!\!,\!\hat{w}_i^{\bar z \bar z'}\!) \!-\! \hat f_i(\hat{x}_i^{\bar z'\!\bar z}\!,\!\hat{w}_i^{\bar z'\!\bar z})\| + 2\mathsf d\big)\\
		&~~+\mathcal S_i^*(q_i,\!f_i(\hat x^{\bar z\bar z'}_i\!\!,\!\hat w^{\bar z\bar z'}_i),\!f_i(\hat x^{\bar z'\!\bar z}_i\!\!,\!\hat w^{\bar z'\!\bar z}_i)\!) - \gamma_i^*\mathcal S_i^*(q_i,\hat x^{\bar z\bar z'}_i\!\!,\hat x^{\bar z'\!\bar z}_i)\\
		&~~-\rho_i^*\Vert  \hat w^{\bar z\bar z'}_i-\hat w^{\bar z'\!\bar z}_i\Vert^2\\
		&= \big(4\mathscr L_i^{\hat f}L_i(2\mathsf d + \mathscr L_i^{\hat f}\varepsilon_i)+ \gamma_i^*\mathscr L_i^{\mathcal S} + {4\sqrt{2}}\rho_i^*\big)\varepsilon_i\\
		&~~+2\mathscr L_i^{\hat f}\varepsilon_iL_i\|\hat f_i(\hat{x}_i^{\bar z \bar z'}\!\!,\!\hat{w}_i^{\bar z \bar z'}\!) \!-\! \hat f_i(\hat{x}_i^{\bar z'\!\bar z}\!,\!\hat{w}_i^{\bar z'\!\bar z})\| \\
		&~~+2\mathsf d L_i\big(\|\hat f_i(\hat{x}_i^{\bar z \bar z'}\!\!,\!\hat{w}_i^{\bar z \bar z'}\!) \!-\! \hat f_i(\hat{x}_i^{\bar z'\!\bar z}\!,\!\hat{w}_i^{\bar z'\!\bar z})\| + 2\mathsf d\big)\\
		&~~+\mathcal S_i^*(q_i,\!f_i(\hat x^{\bar z\bar z'}_i\!\!,\!\hat w^{\bar z\bar z'}_i),\!f_i(\hat x^{\bar z'\!\bar z}_i\!\!,\!\hat w^{\bar z'\!\bar z}_i)\!) - \gamma_i^*\mathcal S_i^*(q_i,\hat x^{\bar z\bar z'}_i\!\!,\hat x^{\bar z'\!\bar z}_i)\\
		&~~-\rho_i^*\Vert  \hat w^{\bar z\bar z'}_i-\hat w^{\bar z'\!\bar z}_i\Vert^2.
	\end{align*}
	Since $\|\hat f_i(\hat{x}_i^{\bar z \bar z'}\!\!,\!\hat{w}_i^{\bar z \bar z'}\!) - \hat f_i(\hat{x}_i^{\bar z'\!\bar z}\!,\!\hat{w}_i^{\bar z'\!\bar z})\|\leq\max_{(z,z')\in\bar{\mathcal N}_i}\|\hat f_i(\hat{x}_i^{zz'}\!\!,$ $\!\hat{w}_i^{zz'}\!) \!-\! \hat f_i(\hat{x}_i^{z'\!z}\!,\!\hat{w}_i^{z'\!z})\|$, we have
	\begin{align*}
		&\mathcal S_i^*(q_i,f_i(x_i,w_i),f_i(x'_i,w'_i)\!) \!-\!\gamma_i^*\mathcal S_i^*(q_i,x_i,x'_i)\!-\!\rho_i^*\Vert w_i\!-\!w'_i\Vert^2\\
		&\leq \big(4\mathscr L_i^{\hat f}L_i(2\mathsf d + \mathscr L_i^{\hat f}\varepsilon_i)+ \gamma_i^*\mathscr L_i^{\mathcal S} + {4\sqrt{2}}\rho_i^*\big)\varepsilon_i\\
		&~~+2\mathscr L_i^{\hat f}\varepsilon_iL_i\max_{\bar{\mathcal N}_i}\|\hat f_i(\hat{x}_i^{zz'}\!\!,\!\hat{w}_i^{zz'}\!) \!-\! \hat f_i(\hat{x}_i^{z'\!z}\!,\!\hat{w}_i^{z'\!z})\| \\
		&~~+2\mathsf d L_i\big(\|\hat f_i(\hat{x}_i^{\bar z \bar z'}\!\!,\!\hat{w}_i^{\bar z \bar z'}\!) \!-\! \hat f_i(\hat{x}_i^{\bar z'\!\bar z}\!,\!\hat{w}_i^{\bar z'\!\bar z})\| + 2\mathsf d\big)\\
		&~~+\mathcal S_i^*(q_i,\!f_i(\hat x^{\bar z\bar z'}_i\!\!,\!\hat w^{\bar z\bar z'}_i),\!f_i(\hat x^{\bar z'\!\bar z}_i\!\!,\!\hat w^{\bar z'\!\bar z}_i)\!) - \gamma_i^*\mathcal S_i^*(q_i,\hat x^{\bar z\bar z'}_i\!\!,\hat x^{\bar z'\!\bar z}_i)\\
		&~~-\rho_i^*\Vert  \hat w^{\bar z\bar z'}_i-\hat w^{\bar z'\!\bar z}_i\Vert^2\\
		&= \big(4\mathscr L_i^{\hat f}L_i(2\mathsf d + \mathscr L_i^{\hat f}\varepsilon_i)+ \gamma_i^*\mathscr L_i^{\mathcal S} + {4\sqrt{2}}\rho_i^*\\
		&~~+2\mathscr L_i^{\hat f}L_i\max_{\bar{\mathcal N}_i}\|\hat f_i(\hat{x}_i^{zz'}\!\!,\!\hat{w}_i^{zz'}\!) \!-\! \hat f_i(\hat{x}_i^{z'\!z}\!,\!\hat{w}_i^{z'\!z})\|\big)\varepsilon_i \\
		&~~+2\mathsf d L_i\big(\|\hat f_i(\hat{x}_i^{\bar z \bar z'}\!\!,\!\hat{w}_i^{\bar z \bar z'}\!) \!-\! \hat f_i(\hat{x}_i^{\bar z'\!\bar z}\!,\!\hat{w}_i^{\bar z'\!\bar z})\| + 2\mathsf d\big)\\
		&~~+\mathcal S_i^*(q_i,\!f_i(\hat x^{\bar z\bar z'}_i\!\!,\!\hat w^{\bar z\bar z'}_i),\!f_i(\hat x^{\bar z'\!\bar z}_i\!\!,\!\hat w^{\bar z'\!\bar z}_i)\!) - \gamma_i^*\mathcal S_i^*(q_i,\hat x^{\bar z\bar z'}_i\!\!,\hat x^{\bar z'\!\bar z}_i)\\
		&~~-\rho_i^*\Vert  \hat w^{\bar z\bar z'}_i-\hat w^{\bar z'\!\bar z}_i\Vert^2\\
		&= \mathscr L_i^3\varepsilon_i +\Psi_i,
	\end{align*}
	where 
	\begin{align*}
		&\Psi_i\coloneq \mathcal S_i^*(q_i,\!f_i(\hat x^{\bar z\bar z'}_i\!\!,\!\hat w^{\bar z\bar z'}_i\!),\!f_i(\hat x^{\bar z'\!\bar z}_i\!\!,\!\hat w^{\bar z'\!\bar z}_i)\!) \!-\! \gamma_i^*\mathcal S_i^*(q_i,\hat x^{\bar z\bar z'}_i\!\!,\hat x^{\bar z'\!\bar z}_i)\\
		&-\rho_i^*\Vert  \hat w^{\bar z\bar z'}_i-\hat w^{\bar z'\!\bar z}_i\Vert^2\\
		&+2\mathsf d L_i\big(\|\hat f_i(\hat{x}_i^{\bar z \bar z'}\!\!,\!\hat{w}_i^{\bar z \bar z'}\!) \!-\! \hat f_i(\hat{x}_i^{\bar z'\!\bar z}\!,\!\hat{w}_i^{\bar z'\!\bar z})\| + 2\mathsf d\big).
	\end{align*}
	Since condition~\eqref{SOP3-1-n} of $\text{SOP}_\varsigma$ is satisfied with $\mu_{\varsigma_{3_i}}^*$, for $\hat L_i\geq L_i$ as per~\eqref{eq:L hat}, one can deduce that $\Psi_i\leq\mu^*_{\varsigma_{3_i}}$. Consequently, we have
\begin{align*}
	&\mathcal S_i^*(q_i,f_i(x_i,w_i),f_i(x'_i,w'_i)\!) \!\!-\!\gamma_i^*\mathcal S_i^*(q_i,x_i,x'_i)\!\!-\!\rho_i^*\Vert w_i\!-\!w'_i\Vert^2\\
	&\leq \mu_{\varsigma_{3_i}}^* + \mathscr L_i^3\varepsilon_i.
\end{align*}}

Given our proposed condition in~\eqref{Con3-3}, one can conclude that
\begin{align}\notag
	&\mathcal S_i^*(q_i,f_i(x_i,w_i),f_i(x'_i,w'_i)) -\gamma_i^*\mathcal S_i^*(q_i,x_i,x'_i)\\\notag
	&-\rho_i^*\Vert w_i\!-\!w'_i\Vert^2\!\leq\! 0,\\\notag
	&\forall {\mathcal Q_i} \in \mathbb R^{n_i}\times \mathbb R^{p_i}\times \mathbb R^{n_i}\times \mathbb R^{p_i}\!\!:\quad
	\Vert{ \mathcal Q_i}\Vert = 1.
\end{align}
{Finally, $\mu_{R_{j_i}} \coloneq \frac{1}{2}(\mu_{\varsigma_{j_i}}^* + \mathscr L_i^j \varepsilon_i)\leq 0$, for $j \in \{1,\dots,3\}$, is a valid choice for satisfying~\eqref{ROP1}–\eqref{ROP3}. In particular, since $x_i$ and $x_i'$ are chosen from the unit sphere $\Vert (x_i,w_i,x'_i,w'_i)\Vert = 1$, it follows that $\Vert x_i - x_i'\Vert \leq \sqrt{2}$. Accordingly, $\mu_{R_{j_i}}$ can be chosen as $\mu_{R_{j_i}} \coloneq \frac{1}{2}(\mu_{\varsigma_{j_i}}^* + \mathscr L_i^j \varepsilon_i),$  since $\mu_{\varsigma_{j_i}}^* + \mathscr L_i^j \varepsilon_i \leq \mu_{R_{j_i}} \Vert x_i - x_i' \Vert^2 \leq 0$ always holds.}

Thus, $\mathcal{S}_i^*$ derived from solving {$\text{SOP}_\varsigma$ in~\eqref{SOP-1-n}} act as $\delta$-ISS functions for unknown subsystems $\Sigma_i, i \!\in\! \{1, \dots, \mathcal M \!\}$, ensuring correctness guarantees, and thereby completing the proof. $\hfill\blacksquare$

Theorem~\ref{Thm2} establishes the $\delta$-ISS Lyapunov functions $\mathcal{S}^*_i$ purely based on data, ensuring that conditions~\eqref{Eq:8_21}–\eqref{Eq:8_22} hold locally on the unit sphere $\Vert(x_i, w_i,x'_i,w'_i)\Vert = 1$. Utilizing the \emph{homogeneity} of both the dynamics $f_i$ and the $\delta$-ISS Lyapunov functions $\mathcal{S}^*_i$, these local conditions are then systematically extended to the entire space $\mathbb{R}^{n_i} \times \mathbb{R}^{p_i}\times \mathbb{R}^{n_i} \times \mathbb{R}^{p_i}$.
\begin{algorithm*}[t]
	\footnotesize
	\caption{Estimation of $\mathscr{L}_i^{\hat f}$ via data {with both asymptotic confidence and quantified confidence level}}
	\label{Alg:1}
	\centering
	
	\begin{minipage}[t]{0.48\textwidth}
		\vspace{0pt}
		\begin{algorithmic}[1]
			\REQUIRE $\underline{\alpha}_i^*, \overline{\alpha}_i^*, \gamma_i^*, \rho_i^*$; 
			$\phi,\sigma \in \mathbb N^+$; $\kappa \in \mathbb R^+$; 
			{mode $m_i \in \{\textsf{conf}_\textsf{lim},\textsf{conf}_\textsf{quan}\}$}
			{\IF{$m_i=\textsf{conf}_\textsf{quan}$}
				\STATE Choose confidence level $\varkappa_i \in (0,1)$ and Kolmogorov–Smirnov (KS) significance level $\vartheta_i^{\mathrm{KS}} \in (0,1)$ (commonly, $\vartheta_i^{\mathrm{KS}} = 0.05$)
				\ENDIF}
			
			\FOR{$k=1\!:\!\sigma$}
			\FOR{$j=1\!:\!\phi$}
			\STATE From $\mathbb R^{n_i}\!\times\!\mathbb R^{p_i}$, select sampled pairs 
			$(\hat{x}_{i}^{zz'},\hat{w}_{i}^{zz'})^{k,j}$ and $(\hat{x}_{i}^{'zz'},\hat{w}_{i}^{'zz'})^{k,j}$
			such that
			\[
			\bigl\|(\hat{x}_{i}^{zz'},\hat{w}_{i}^{zz'})^{k,j}-(\hat{x}_{i}^{'zz'},\hat{w}_{i}^{'zz'})^{k,j}\bigr\|
			\le \kappa
			\]
			\STATE Compute the slope
			\[
			\theta_{k,j}
			=
			\frac{
				\bigl\|
				\hat f_i(\hat{x}_{i}^{zz'},\hat{w}_{i}^{zz'})^{k,j}
				-
				\hat f_i(\hat{x}_{i}^{'zz'},\hat{w}_{i}^{'zz'})^{k,j}
				\bigr\|
			}{
				\bigl\|
				(\hat{x}_{i}^{zz'},\hat{w}_{i}^{zz'})^{k,j}
				-
				(\hat{x}_{i}^{'zz'},\hat{w}_{i}^{'zz'})^{k,j}
				\bigr\|
			}
			\]
			\ENDFOR
			\STATE Set the block maximum
			\[
			\Theta_k=\max\{\theta_{k,1},\dots,\theta_{k,\phi}\}
			\]
			\ENDFOR
			
			\STATE Fit a Reverse Weibull distribution to $\Theta_1,\dots,\Theta_\sigma$ by maximum likelihood and obtain its location, scale, and shape parameters, {denoted by $(\hat a_i,\hat b_i,\hat c_i)$}
			\STATE Set $\mathscr{L}_{i}^{\hat f}=\hat a_i$
		\end{algorithmic}
	\end{minipage}
	\hfil
	\begin{minipage}[t]{0.48\textwidth}
		\vspace{0pt}
		\begin{algorithmic}[1]
			\makeatletter
			\setcounter{ALG@line}{12}
			\makeatother
			{\IF{$m_i=\textsf{conf}_\textsf{lim}$}
				\STATE \textbf{return} $\mathscr{L}_{i}^{\hat f}$
				\ELSE
				\STATE Perform a one-sample KS goodness-of-fit test for\vspace{-0.15cm}
				\[
				\begin{aligned}
					&H_0\!:\ \Theta_1,\dots,\Theta_\sigma
					\text{ follow the fitted Reverse Weibull}\\ &\hphantom{H_0\!:~}\text{ distribution}
				\end{aligned}\vspace{-0.15cm}
				\]
				and let $\pi_i^{\mathrm{KS}}$ denote the associated $p$-value
				\IF{$\pi_i^{\mathrm{KS}} \le \vartheta_i^{\mathrm{KS}}$}
				\STATE Increase $\phi$ and/or $\sigma$, or adjust $\kappa$, and repeat Steps $4$--$10$
				\ELSE
				\STATE Compute the standard error $\aleph_i$ of the fitted location parameter $\hat a_i$
				\STATE Compute the {overapproximation} radius
				$$
				\mathcal C_i=\Phi^{-1}(\varkappa_i)\,\aleph_i
				$$
				where $\Phi$ is the cumulative distribution function of the standard normal distribution
				\STATE Set the confidence-bound estimate
				$$
				\mathscr{L}_{i}^{\hat f}\gets\mathscr{L}_{i}^{\hat f}+\mathcal C_i
				$$
				\STATE \textbf{return} $\mathscr{L}_{i}^{\hat f}$
				\ENDIF
				\ENDIF}
			
			\ENSURE $\mathscr{L}_{i}^{\hat f}$, {$\mathcal C_i$ if $m_i=\textsf{conf}_\textsf{quan}$}
		\end{algorithmic}
	\end{minipage}
	
\end{algorithm*}
{\begin{remark}\label{New1}
		We consider the measurement noise to vary smoothly with the state. Nevertheless, this still corresponds to measurement noise, not process noise, since the perturbation only affects the observed or measured dynamics and does not alter the true underlying system dynamics. In other words, the system itself evolves according to the true dynamics, while the noise only enters through the sensing or estimation process used to obtain the measured map. This situation is consistent with many practical sensing and estimation pipelines where the effective measurement error varies smoothly with the operating point. For instance, sensor bias or calibration errors can depend on the state, and interpolation or regression applied to measurement data can also introduce smooth approximation errors. As another practical scenario, in autonomous vehicles, velocity or pose estimates obtained from cameras or IMU sensors can exhibit errors that vary smoothly with the vehicle’s operating state due to calibration biases or the geometry of the sensing setup. In addition, observer-based or estimator-based methods used to process measurements can introduce errors that depend on the operating state.
		
		We note that even when the underlying sensor signal is initially non-smooth, the measured signals typically pass through filtering, smoothing, or estimation stages. These stages often produce effective measurement errors that vary continuously with respect to the state. Under these considerations, the measured dynamics map $\hat f_i(\cdot,\cdot)$, as defined in~\eqref{eq:noisy measurements}, is also Lipschitz continuous since $f_i(\cdot,\cdot)$ is Lipschitz continuous. We highlight that for different cases with various measurement noise in practice, if the estimated Lipschitz constant for $\hat f_i(\cdot,\cdot)$ remains bounded over sufficiently large samples, it provides empirical evidence that the noise-corrupted $\hat f_i(\cdot,\cdot)$ behaves Lipschitz over the region of interest, which is the unit sphere in our setting. \hfill $\square$
\end{remark}}

\subsection{Lipschitz Constant Estimation from Data}\label{Lipschitz}

To verify conditions~\eqref{Con3-1}--\eqref{Con3-3} in Theorem~\ref{Thm2}, computing {$\mathscr L_i^1$, $\mathscr L_i^2$, and $\mathscr{L}_{i}^3$, as defined in~\eqref{eq:L3}, are necessary steps.} For this purpose, we propose Algorithm~\ref{Alg:1}, which facilitates the estimation of {$\mathscr{L}_i^{\hat f}$} required for the computation of {$\mathscr{L}_{i}^3$}, using a finite set of data. {This algorithm consists of two components: one, corresponding to $m_i = \textsf{conf}_\textsf{lim}$, guarantees asymptotic correctness in the limit and is adopted from~\citep{wood1996estimation}; the other, corresponding to $m_i = \textsf{conf}_\textsf{quan}$, quantifies an {overapproximation radius} and yields an overestimation of the Lipschitz constant, adopted from~\citep{weng2018evaluating, knuth2021planning}.} The first part of Algorithm~\ref{Alg:1} relies on Lemma~\ref{lemma} to ensure the convergence of the estimated {$\mathscr{L}_i^{\hat f}$} towards its true value within the limit~\citep{wood1996estimation}.
\begin{lemma}\label{lemma}
	In Algorithm~\ref{Alg:1}, the estimated {$\mathscr{L}_i^{\hat f}$} converges to its actual value if and only if $\kappa$ approaches zero, while both $\phi$ and $\sigma$ tend to infinity.
\end{lemma}
{According to Lemma~\ref{lemma}, the convergence of the Lipschitz constant holds in the limit, and the estimated Lipschitz constant using a limited amount of data is typically associated with a confidence bound, which can be quantified when using the second part of Algorithm~\ref{Alg:1} with $m_i=\textsf{conf}_\textsf{quan}$.}

{\begin{remark}\label{rem:lip analytic}
		We note that $\bar{\mathcal S}_i^*(e_i)\coloneq\mathcal S_i^*(q_i,x_i,x'_i)$, where $e_i\coloneq x_i-x'_i$, is Lipschitz continuous with respect to $(x_i,x'_i)$ over the compact unit sphere set $\|(x_i,w_i,x'_i,w'_i)\|=1$, with Lipschitz constant $\mathscr L_i^{\mathcal S}$. Once the $\delta$-ISS Lyapunov function is obtained via solving $\text{SOP}_{\varsigma}$~\eqref{SOP-1-n} in our data-driven framework, one can compute $\mathscr L_i^{\mathcal S}$ analytically. Similarly, functions $\underline\Upsilon_i^*(e_i)\coloneq\underline{\alpha}_i^*{\Vert e_i\Vert^2} - \bar{\mathcal S}_i^*(e_i)$ and $\overline\Upsilon_i^*(e_i)\coloneq\bar{\mathcal S}_i^*(e_i) - \overline{\alpha}_i^*{\Vert e_i\Vert^2}$ are both Lipschitz continuous with respect to $(x_i,x'_i)$, with Lipschitz constants $\mathscr{L}^1_i$ and $\mathscr{L}^2_i$, respectively, and once their corresponding parameters are determined, one can compute these Lipschitz constants analytically, as well. 
		
		To do so, considering the compact unit sphere set $\|(x_i,w_i,x'_i,w'_i)\|=1$, resulting in $\mathcal E_i\coloneq\{e_i:\|e_i\|\leq\sqrt{2}\}$, one has
		\begin{align*}
			\mathscr L_i^{\mathcal S} &= \max_{\mathcal E_i} \big\Vert \nabla \bar{\mathcal S}_i^*(e_i) \big\Vert,\;
			\mathscr L_i^{1} = \max_{\mathcal E_i} \big\Vert \nabla \underline\Upsilon_i^*(e_i) \big\Vert,\\
			\mathscr L_i^{2} &= \max_{\mathcal E_i} \big\Vert \nabla \overline\Upsilon_i^*(e_i) \big\Vert.
		\end{align*}
		As a result, among the four Lipschitz constants appearing in~\eqref{eq:rob cons}, three can be derived analytically, whereas only one, namely $\mathscr L_i^{\hat f}$, should be estimated from data. Therefore, assuming that only this single Lipschitz constant is {provided as a valid upper bound or conservatively estimated from data} is not overly restrictive. \hfill $\square$
\end{remark}}
{\begin{remark}
		To provide an intuitive illustration of the relationship between the user-specified confidence level $\varkappa_i$ and the resulting {overapproximation} radius $\mathcal C_i$ obtained from the second part of Algorithm~\ref{Alg:1}, i.e., by setting $m_i = \textsf{conf}_\textsf{quan}$, we present the following example. Let $\aleph_i = 1$ in Step~21 of Algorithm~\ref{Alg:1}, which results in $\mathcal{C}_i = \Phi^{-1}(\varkappa_i)$. Consequently, for $\varkappa_i = 0.9$, we obtain
		$$
		\mathcal C_i = \Phi^{-1}(0.9)\approx 1.2816,
		$$
		for $\varkappa_i = 0.99999$, we obtain
		$$
		\mathcal C_i = \Phi^{-1}(0.99999)\approx 4.2649,
		$$
		and for $\varkappa_i = 1$, we have
		$$
		\mathcal C_i = \Phi^{-1}(1) \approx +\infty,
		$$
		since the cumulative distribution function of the standard normal distribution approaches $1$ asymptotically and does not attain it at any finite argument. This example illustrates that, as the prescribed confidence level $\varkappa_i$ increases, the {overapproximation} radius $\mathcal{C}_i$ also increases, leading to a more conservative estimate of the Lipschitz constant $\mathscr{L}_i^{\hat f}$, as expected. Nevertheless, the role of the standard error $\aleph_i$ in the fitted location parameter (set to $1$ in this example) in Step~20 of Algorithm~\ref{Alg:1} is equally important, as smaller values of this error can still moderate the {overapproximation} radius $\mathcal{C}_i$ even for high confidence levels $\varkappa_i$. We highlight that collecting more data for estimating the Lipschitz constant can potentially reduce the value of the standard error $\aleph_i$. \hfill $\square$
\end{remark}}
In the following section, we study the interconnected network and introduce a compositional framework based on small-gain reasoning for developing an incremental GAS certificate for the interconnected network based on $\delta$-ISS Lyapunov functions $\mathcal S_i$ for individual subsystems, which are derived from data.

\section{Incremental GAS Certificate for Interconnected Network}\label{Guarantee_ROP}

Prior to introducing the primary compositionality result of this work, we define $ \Gamma \coloneq \mathsf{diag}(\hat\gamma_1, \ldots, \hat\gamma_{\mathcal M})$ with $\hat\gamma_i = 1 - \gamma_i^*$, and $\Delta \coloneq \{\hat\delta_{ij}\}$ where $\hat\delta_{ij} = \frac{\rho_i^*}{\underline\alpha_j^*}$ and $\hat\delta_{ii} = 0$ for all $i \in\{1, \ldots, \mathcal M\}$. {Additionally, we assume $\hat \delta_{ij} = 0$ when there is no interconnection from $\Sigma_j$ to $\Sigma_i$.}

In the upcoming theorem, we offer our results for constructing an incremental GAS certificate for the interconnected network using $\delta$-ISS Lyapunov functions $\mathcal S_i^*$ for individual subsystems, obtained from data.

\begin{theorem}\label{Thm:3}
	Consider an interconnected dt-NS $\Sigma = \mathcal{I}(\Sigma_1,\ldots,\Sigma_{\mathcal M})$, comprising $\mathcal M\in\mathbb{N}^+$ individual subsystems~$\Sigma_i$ with unknown dynamics.  Suppose that each $\Sigma_i$ admits a data-driven $\delta$-ISS Lyapunov function $\mathcal S_i^*$ with a correctness guarantee, as proposed in Theorem~\ref{Thm2}. If 
	\begin{align}\notag
		&\mathds{1}_{\mathcal M}^\top(-\Gamma+\Delta) \coloneq \big[\zeta_1;\dots;\zeta_{\mathcal M}\big]^\top < 0, \\\label{Eq:43}
		&\text{equivalently}, ~ \zeta_i < 0, \forall i\in\{1,\dots,\mathcal M\},
	\end{align} 
	then 
	\begin{align}\label{Lyp}
		\mathcal V(q,x,x') \coloneq\ \sum_{i=1}^{\mathcal M}\mathcal S_i^*(q_i,x_i,x_i')
	\end{align}
	is an incremental Lyapunov function for the unknown interconnected dt-NS $\Sigma = \mathcal{I}(\Sigma_1,\ldots,\Sigma_{\mathcal M})$ with
	\begin{align*}
		&\quad\quad\quad\quad\underline{\alpha} = \underset{i}{\min}\{\underline{\alpha}_i^*\},\quad \overline{\alpha}=\underset{i}{\max}\{\overline{\alpha}_i^*\},\\
		&\gamma\coloneq 1 +\zeta, ~ \text{where} ~  \max_{1\leq i\leq \mathcal M} \zeta_i < \zeta < 0, ~ \text{and} ~ \zeta \in (-1,0).
	\end{align*}
	Accordingly, $\Sigma = \mathcal{I}(\Sigma_1,\ldots,\Sigma_{\mathcal M})$ is $\delta$-GAS in the sense of Theorem~\ref{GA1S}. 
\end{theorem}
\begin{figure*}[t!]
	\rule{\textwidth}{0.1pt}
	\begin{align}\notag
		&\mathcal V(q,f(x), f(x'))\!=\!\sum_{i=1}^{\mathcal M}\!\mathcal S_{i}^*(q_i,f_i(x_i,w_i), f_i(x_i',w_i'))\\\notag
		&\leq\sum_{i=1}^{\mathcal M}\big(\gamma_i^*\mathcal S_i^*(q_i,x_i,x'_i)+\rho_i^*\Vert w_i-w'_i\Vert^2\big)\leq\sum_{i=1}^{\mathcal M}\big(\gamma_i^*\mathcal S_i^*(q_i,x_i,x'_i)\!+\!\rho_{i}^*\sum_{\substack{j=1\\i\neq{j}}}^{\mathcal M}\Vert w_{ij}-w'_{ij}\Vert^2\big)\\\notag
		&= \sum_{i=1}^{\mathcal M}\big(\gamma_i^*\mathcal S_i^*(q_i,x_i,x'_i)\!+\!\rho_{i}^*\sum_{\substack{j=1\\i\neq{j}}}^{\mathcal M}\Vert x_{j} - x_j'\Vert^2\big) =\sum_{i=1}^{\mathcal M}\big(\gamma_i^*\mathcal S_i^*(q_i,x_i,x'_i)\!+\!\sum_{\substack{j=1\\i\neq{j}}}^{\mathcal M}\rho_{i}^*\Vert x_{j} - x_j'\Vert^2\big)\\\notag
		&\!\leq\!\sum_{i=1}^{\mathcal M}\big(\gamma_i^*\mathcal S_i^*(q_i,x_i,x'_i)\!+\!\sum_{\substack{j=1\\i\neq{j}}}^{\mathcal M}\frac{\rho_{i}^*}{\underline\alpha_{j}^*}\mathcal S_{j}^*(q_j,x_j,x_j')\big)\!=\!\sum_{i=1}^{\mathcal M}\mathcal S_{i}^* (q_i,x_i,x_i'){+\! \sum_{i=1}^{\mathcal M}\big(-}{\hat\gamma_i}\mathcal S_i^*(q_i, x_i,x_i')+\!\sum_{\substack{j=1\\i\neq{j}}}^{\mathcal M}\hat\delta_{ij}\mathcal S_{j}^*(q_j,x_j,x_j')\big)\\\label{Eq:17}
		&=\mathcal V(q,x,x') + \mathds{1}_{\mathcal M}^\top(-\Gamma+\Delta)\big[\mathcal S_1^*(q_1,x_1,x_1');\ldots;\mathcal S_{\mathcal M}^*(q_{\mathcal M},x_{\mathcal M},x_{\mathcal M}')\big]\leq \underbrace{(1 + \zeta)}_{\gamma} \mathcal V(q,x,x').
	\end{align}
	\rule{\textwidth}{0.1pt}
\end{figure*}
{\bf Proof.}  We initiate with showing the lower bound in condition~\eqref{alpha12}. Given the proposed form of the incremental Lyapunov function in~\eqref{Lyp} and {using the definition of the Euclidean norm for stacked vectors}, one has:
\begin{align*}
	&\underline{\alpha}\Vert x\!-\!x'\Vert^2 \!-\! \mathcal V(q,x,x') \!\overset{\eqref{Lyp}}{=}\! \underline{\alpha}\Vert x\!-\!x'\Vert^2 \!-\! \sum_{i=1}^{\mathcal M}  \mathcal S_i^*(q_i,x_i,x'_i)\\
	& = \underline{\alpha}\Vert [x_1;\dots;x_{\mathcal M}]\!-\![x'_1;\dots;x'_{\mathcal M}]\Vert^2 \!-\! \sum_{i=1}^{\mathcal M}  \mathcal S_i^*(q_i,x_i,x'_i)\\
	&= \underline{\alpha}\sum_{i=1}^{\mathcal M} \Vert x_i-x'_i\Vert^2 - \sum_{i=1}^{\mathcal M}  \mathcal S_i^*(q_i,x_i,x'_i)\\ 
	&= \sum_{i=1}^{\mathcal M} \big(\underline{\alpha}\Vert x_i-x'_i\Vert^2 - \mathcal S_i^*(q_i,x_i,x'_i)\big).
\end{align*}
By considering  $\underline{\alpha} = \underset{i}\min\{\underline{\alpha}_i^*\}$, one has
\begin{align}\notag
	&\underline{\alpha}\Vert x-x'\Vert^2 - \mathcal V(q,x,x') \\\notag
	&= \sum_{i=1}^{\mathcal M} \big(\min_i\{\underline{\alpha}_i^*\}\Vert x_i-x'_i\Vert^2 - \mathcal S_i^*(q_i,x_i,x'_i)\big)\\\label{new}
	&\leq \sum_{i=1}^{\mathcal M} \big(\underline{\alpha}_i^*\Vert x_i-x'_i\Vert^2 - \mathcal S_i^*(q_i,x_i,x'_i)\big).
\end{align}
According to the lower bound of condition~\eqref{Eq:8_21}, one has $\underline{\alpha}_i^*\Vert x_i-x'_i\Vert^2 - \mathcal S_i^*(q_i,x_i,x'_i) \leq 0$. Then according to \eqref{new}, we have
\begin{align*}
	\underline{\alpha} \Vert x-x' \Vert^2 \leq \mathcal V(q,x,x').
\end{align*}

Under a similar argument with $\overline{\alpha} = \underset{i}\max\{\overline{\alpha}_i^*\}$ and under the upper bound of condition~\eqref{Eq:8_21}, one can show that
\begin{align*}
	&\mathcal V(q,x,x') - \overline{\alpha}\Vert x-x'\Vert^2   \overset{\eqref{Lyp}}{=} \sum_{i=1}^{\mathcal M}  \mathcal S_i^*(q_i,x_i,x'_i) - \overline{\alpha}\Vert x-x'\Vert^2 \\
	& =\sum_{i=1}^{\mathcal M}  \mathcal S_i^*(q_i,x_i,x'_i) - \overline{\alpha}\Vert [x_1;\dots;x_{\mathcal M}]\!-\![x'_1;\dots;x'_{\mathcal M}]\Vert^2 \\
	& = \sum_{i=1}^{\mathcal M}  \mathcal S_i^*(q_i,x_i,x'_i) - \overline{\alpha}\sum_{i=1}^{\mathcal M} \Vert x_i-x'_i\Vert^2\\ 
	&= \sum_{i=1}^{\mathcal M} \big(\mathcal S_i^*(q_i,x_i,x'_i) - \overline{\alpha}\Vert x_i-x'_i\Vert^2 \big)\\
	&= \sum_{i=1}^{\mathcal M} \big( \mathcal S_i^*(q_i,x_i,x'_i)  - \max_i\{\overline{\alpha}_i^*\}\Vert x_i-x'_i\Vert^2 \big)\\
	&\leq \sum_{i=1}^{\mathcal M} \big(\mathcal S_i^*(q_i,x_i,x'_i) - \overline{\alpha}_i^*\Vert x_i-x'_i\Vert^2 \big) \leq 0,
\end{align*}
which accordingly results in
\begin{align*}
	\mathcal V(q,x,x')\leq\overline{\alpha} \Vert x-x' \Vert^2 .
\end{align*}
We now proceed with showing that under condition~\eqref{Eq:43}, $\mathcal V$ satisfies condition~\eqref{alpha1} as well.  By utilizing the lower bound of condition~\eqref{Eq:8_21} and compositionality condition $\mathds{1}_{\mathcal M}^\top(-\Gamma+\Delta)< 0$, one can obtain the chain of inequalities in \eqref{Eq:17}. By defining
\begin{align}\notag
	&\gamma s \coloneq \max\Big\{s + \mathds{1}_{\mathcal M}^\top(-\Gamma+\Delta)\bar {\mathcal S}(q,x,x')\\\label{Eq:18}
	&\hphantom{\gamma s \coloneq \max \Big\{s }\big|\, \mathds{1}_{\mathcal M}^\top\bar {\mathcal S}(q,x,x')=s\Big\},
\end{align}
where $\bar {\mathcal S}(q,x,x')\coloneq\big[\mathcal S_1^*(q_1,x_1,x_1');\ldots;\mathcal S_{\mathcal M}^*(q_{\mathcal M},x_{\mathcal M},x_{\mathcal M}')\big]$, condition \eqref{alpha1} is also satisfied. 

As the last step of the proof, we now show that $ \gamma = 1 + \zeta$ and $0<\gamma<1$. Since $\mathds{1}_{\mathcal M}^\top(-\Gamma+\Delta) \coloneq \big[\zeta_1;\dots;\zeta_{\mathcal M}\big]^\top < 0$, and  $\max_{1\leq i\leq {\mathcal M}} \zeta_i < \zeta < 0$ with $\zeta \in (-1,0)$, one has
\begin{align*}
	&\gamma s =s \!+\! \mathds{1}_{\mathcal M}^\top(-\Gamma\!+\!\Delta)\bar {\mathcal S}(q,x,x') \\
	&= s \!+\!\! \big[\zeta_1;\dots;\zeta_{\mathcal M}\big]^\top\!\big[\mathcal S_1^*(q_1,\!x_1,\!x_1');
	\ldots;\mathcal S_{\mathcal M}^*(q_{\mathcal M},\!x_{\mathcal M},\!x_{\mathcal M}')\!\big]\\
	&= s \!+\! \zeta_1\mathcal S_1^*(q_1,x_1,x_1') \!+\! \dots \!+\! \zeta_{\mathcal M}\mathcal S_{\mathcal M}^*(q_{\mathcal M},x_{\mathcal M},x_{\mathcal M}')\\
	& \leq s \!+\! \zeta\big(\mathcal S_1^*(q_1,x_1,x_1') \!+\! \dots \!+\! \mathcal S_{\mathcal M}^*(q_{\mathcal M},x_{\mathcal M},x_{\mathcal M}')\big) \\
	& = s \!+\! \zeta s \!=\! (1 + \zeta) s.
\end{align*}
Hence, $\gamma s \leq (1 + \zeta) s$, and accordingly, $\gamma \leq1 + \zeta$. Since $\max_{1\leq i\leq {\mathcal M}} \zeta_i < \zeta < 0$ with $\zeta \in (-1,0)$, then $0 <\gamma = 1 + \zeta <1$. Thereby, $\mathcal V$ in the form of~\eqref{Lyp} is an incremental Lyapunov function for the unknown interconnected dt-NS $\Sigma = \mathcal{I}(\Sigma_1,\ldots,\Sigma_{\mathcal M})$, ensuring that $\Sigma = \mathcal{I}(\Sigma_1,\ldots,\Sigma_{\mathcal M})$ is $\delta$-GAS, which concludes the proof. $\hfill\blacksquare$

\begin{remark}
	It should be noted that $\hat\gamma_i$ and $\hat\delta_{ij}$ capture the gain of each individual subsystem and its interaction gain with other subsystems in the interconnected topology, specifically $\gamma_{i}$ and $\rho_{i}$, respectively. These $\hat\gamma_i$ and $\hat\delta_{ij}$ are subsequently utilized to construct $\Gamma$ and $\Delta$, thereby establishing the compositionality condition in~\eqref{Eq:43}.\hfill $\square$
\end{remark}

We now propose Algorithm~\ref{Alg:2} to describe the required steps for ensuring the $\delta$-GAS certificate over unknown interconnected dt-NS.

{\begin{remark}
		If conditions~\eqref{Con3-1}--\eqref{Con3-3} are not satisfied, then the proposed procedure cannot certify that the subsystem is $\delta$-ISS. However, this should not be interpreted as evidence that the subsystem fails to be $\delta$-ISS; rather, it only indicates that no certificate has been found with the current choice of basis functions and available data. We note that this limitation is not specific to the present data-driven framework, but is inherent to Lyapunov-based analysis of nonlinear systems, where the success of the verification step depends on the choice of the candidate function. In practice, the likelihood of success can be improved by enriching the set of basis functions and by increasing the number and coverage of data samples, as mentioned in Step~7 of Algorithm~\ref{Alg:2}.  \hfill $\square$
\end{remark}}

\begin{algorithm}[t!]
	\caption{$\delta$-GAS certificate of interconnected dt-NS with unknown dynamics {and noisy data}}
	\label{Alg:2}		
	\begin{center}
		\begin{algorithmic}[1]
			\REQUIRE Collect data-points 
			\begin{align*}
				&( \tilde{x}_i^z, \tilde{w}_i^z,\tilde{x}_i^{z'}\!, \tilde{w}_i^{z'}\!, {\hat f_i( \tilde{x}_i^z, \tilde{w}_i^z),
					\hat f_i (\tilde{x}_i^{z'}, \tilde{w}_i^{z'})\!)},~\forall (z,z') \in \bar{\mathcal{N}_i}
			\end{align*}
			\STATE Project all data-points onto unit sphere by normalizing them as in \eqref{New77}
			\STATE Compute $\varepsilon_{i}$ as the maximum distance between any points on the unit sphere and the set of data points as in \eqref{EQ:41}
			\STATE Solve {$\text{SOP}_\varsigma$ in~\eqref{SOP-1-n}} with the normalized data and obtain {$\mu^*_{\varsigma_{1_i}}$, $\mu^*_{\varsigma_{2_i}}$, and $\mu^*_{\varsigma_{3_i}}$}
			\STATE {Compute Lipschitz constant $\mathscr{L}_{i}^{\mathcal S},\mathscr L_i^1,\mathscr L_i^2$ analytically, and $\mathscr L_i^{\hat f}$ according to Algorithm~\ref{Alg:1}}  
			\STATE If {{$\mu^*_{\varsigma_{j_i}} + \mathscr L_i^j\varepsilon_i  \leq 0$, for $j\in\{1,\dots,3\}$,} as stated in~\eqref{Con3-1}--\eqref{Con3-3}, respectively, then the constructed $\mathcal{S}_i^*$ obtained by solving $\text{SOP}_\varsigma$~\eqref{SOP-1-n}} serve as $\delta$-ISS functions for unknown subsystems $\Sigma_i$ with correctness guarantees
			\STATE If $\mathds{1}_{\mathcal M}^\top(-\Gamma+\Delta) \coloneq \big[\zeta_1;\dots;\zeta_{\mathcal M}\big]^\top < 0$ according to~\eqref{Eq:43}, then unknown interconnected dt-NS is $\delta$-GAS with the incremental Lyapunov function $\mathcal V(q,x,x') \coloneq \sum_{i=1}^{\mathcal M}\mathcal S_i^*(q_i,x_i,x'_i)$
			\STATE Otherwise, repeat Steps 2--5 with more collected data $\bar{\mathcal N}_i$ to potentially reduce $\varepsilon_{i}$, and design other interaction gains $\rho_i^*$ and optimal values {$\mu^*_{\varsigma_{1_i}}$\!, $\mu^*_{\varsigma_{2_i}}$\!, and $\mu^*_{\varsigma_{3_i}}$}\!, possibly {fulfilling conditions~\eqref{Con3-1}--\eqref{Con3-3} or condition~\eqref{Eq:43}}
			{\ENSURE $\mathcal V(q,x,x')$, $\delta$-GAS certificate}
		\end{algorithmic}
	\end{center}
\end{algorithm}

{\section{Case Study: Controlled Duffing Oscillator Network}\label{Case_Study}
	In this section, we apply our data-driven findings to a physical network comprising $10000$ Duffing oscillator subsystems, where each subsystem under the given controller is a nonlinear homogeneous system of degree one, and thereby, highlight the applicability of our results in real-world, large-scale networks with unknown dynamics.
	
	We consider an interconnected network of Duffing oscillators, adapted from~\citep{tellez2022data}, comprising $\mathcal M = 10000$ subsystems with unknown dynamics, where each open-loop subsystem $\Sigma_i^{\mathrm{ol}}$ is described by
	{\begin{align*}
		\Sigma_i^{\mathrm{ol}}\!\!:\!\left\{\!\begin{array}{ll}
			x_{1_i}\!(k\!+\!1) =  &\! x_{1_i}\!(k) + 0.01 x_{2_i}\!(k) + 0.01 u_{1_i}(k),\\
			x_{2_i}\!(k\!+\!1) = &\!0.99 x_{2_i}\!(k) + 0.003 x_{1_i}\!(k)-0.01 x_{1_i}^3\!(k)\\
			{} & + 0.001 w_i(k) + 0.03 u_{2_i}(k),
		\end{array}\right.
	\end{align*}}
	where $w_i = x_{1_{i-1}}$, with $x_{1_0} = x_{1_\mathcal M}$, and {$u_{1_i}$ and $u_{2_i}$ denote the control input, which are given as
	\begin{align*}
		u_{1_i} &= -97x_{1_i},\\
		u_{2_i} &= \frac{1}{3}\big(x_{1_i}^3 -96 x_{2_i} -1.3x_{1_i}+\sqrt{x_{1_i}^2 + x_{2_i}^2}\big).
	\end{align*}}
	Under this control input, the closed-loop subsystem dynamics become
	{\begin{align*}
		\Sigma_i\!\!:\!\left\{\!\begin{array}{ll}
			x_{1_i}\!(k\!+\!1) = &\! 0.03x_{1_i}\!(k) + 0.01 x_{2_i}\!(k),\\
			x_{2_i}\!(k\!+\!1) = &\!0.03 x_{2_i}\!(k) - 0.01 x_{1_i}\!(k)\\
			{} & \!+ 0.01 \sqrt{x_{1_i}^2\!(k) + x_{2_i}^2\!(k)} + 0.001 w_i(k),
		\end{array}\right.
	\end{align*}}
	which is a nonlinear homogeneous system of degree one. Since the dynamics are assumed to be unknown, we verify the degree-one homogeneity property of the network using the approach in Remark~\ref{rem:hom}.}

\begin{figure}[t!]
	\centering
	\subfloat[\centering {A pair of $x_1$ trajectories from an arbitrary subsystem}\label{fig:x1}]{
		\includegraphics[width=0.8\linewidth]{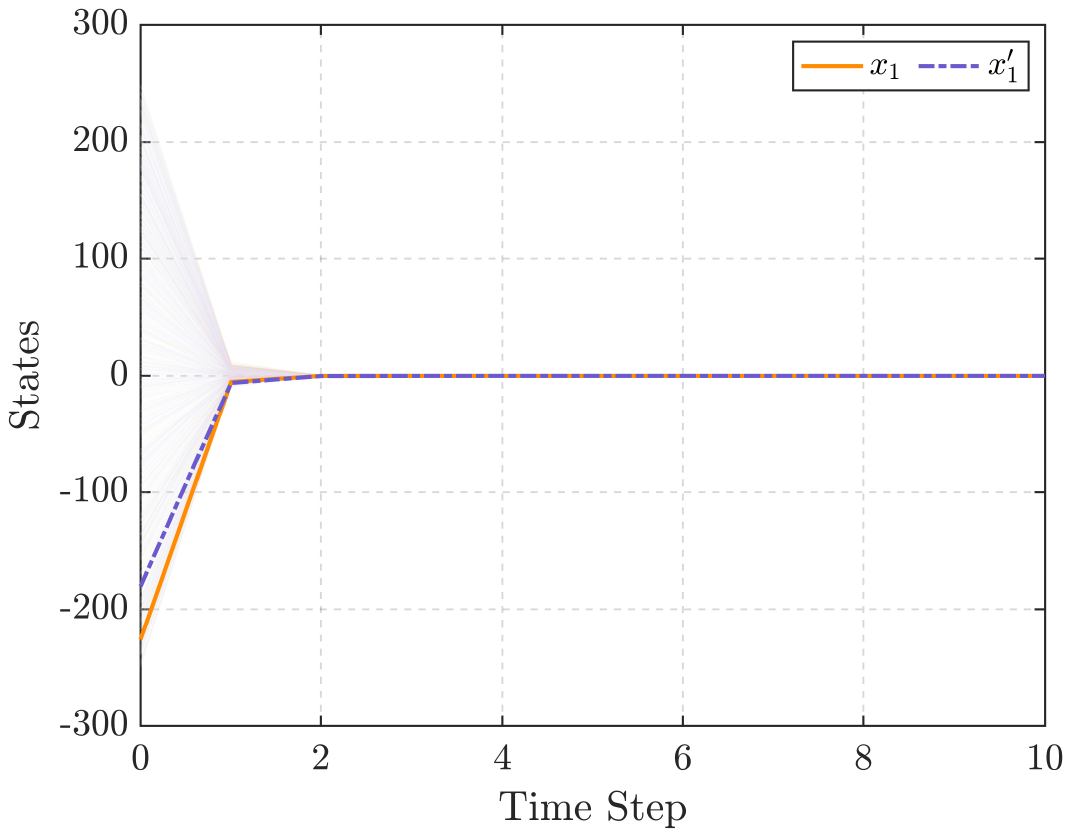}}
	\\
	\subfloat[\centering {A pair of $x_2$ trajectories from an arbitrary subsystem}\label{fig:x2}]{
		\includegraphics[width=0.8\linewidth]{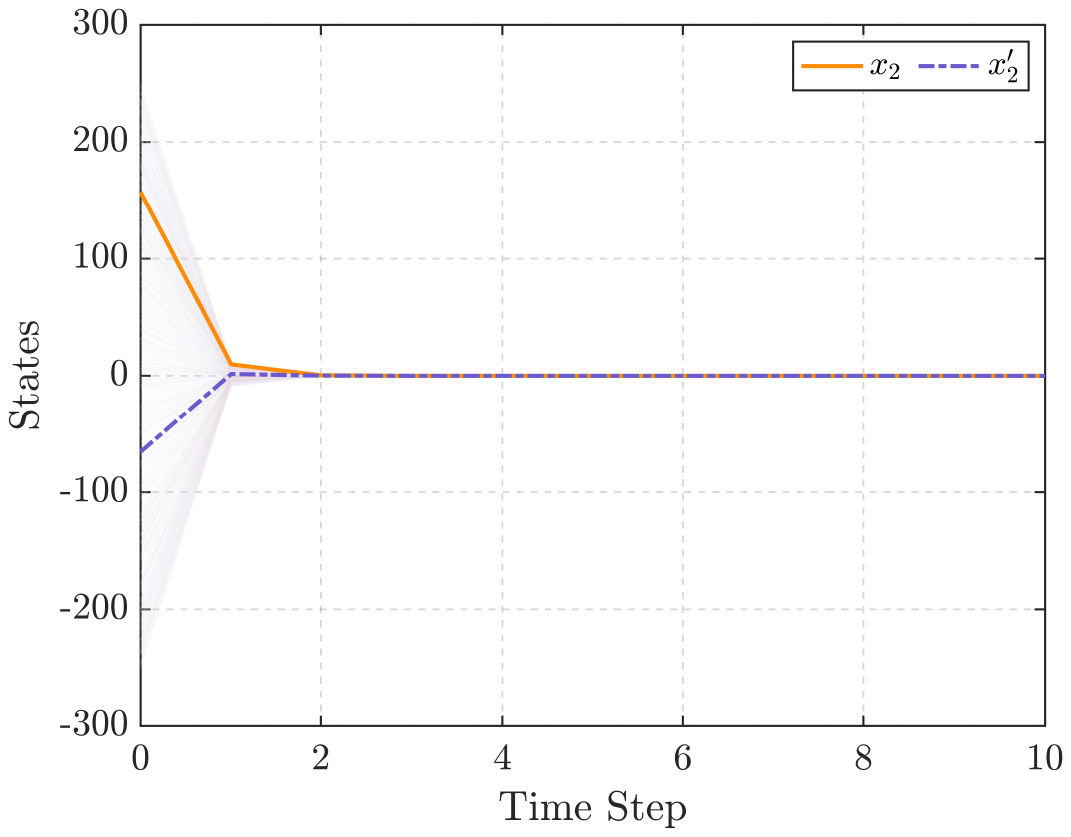}}
	\caption{{{\bf Controlled Duffing oscillator network:} Pairs of $x_1$ and $x_2$ trajectories of all subsystems are demonstrated in faded colors while those of an arbitrary subsystem are depicted in bold colors in Fig. (a) and Fig. (b), respectively.}}
	\label{fig:nonlinear}
\end{figure}
Our primary aim is to construct an \emph{incremental} Lyapunov function for the unknown interconnected dt-NS by utilizing data-driven $\delta$-ISS functions of individual subsystems through solving  {$\text{SOP}_\varsigma$ outlined in~\eqref{SOP-1-n}}. Accordingly, we validate that the interconnected network achieves $\delta$-GAS concerning its equilibrium point $x = 0$, ensuring a correctness guarantee.

Initially, we define the structure of our $\delta$-ISS Lyapunov functions as $\mathcal S_i(q_i,x_i,x_i') = q_{i}^1(x_{1_i}-x_{1_i}')^2 + q_i^2(x_{1_i}-x_{1_i}')(x_{2_i}-x_{2_i}') + q_i^3(x_{2_i} - x_{2_i}')^2$ for all $i\in\{1,\ldots,10000\}$, {and compute $\bar L_i^1 = 2$, $\bar L_i^2 = 1$, and $\bar L_i^3 = 2$, according to Example~\ref{example}}. Our proposed data-driven scheme is implemented using Algorithm~\ref{Alg:2}. We gather {$100000$} sample {tuples} from trajectories of each unknown subsystem and normalize them to project onto the unit sphere. Subsequently, we solve {$\text{SOP}_\varsigma$ in~\eqref{SOP-1-n}}. {Since the $\delta$-ISS Lyapunov candidate is quadratic, conditions~\eqref{SOP1-1-n} and \eqref{SOP2-1-n} are naturally satisfied with $\underline{\alpha}_i^* = \lambda_{\min}(P_i)$ and $\overline{\alpha}_i^* = \lambda_{\max}(P_i)$ once a positive-definite matrix $P_i$, containing the $\delta$-ISS Lyapunov coefficients, is obtained by enforcing condition~\eqref{SOP3-1-n}. Therefore, we restrict our attention to satisfying condition~\eqref{SOP3-1-n} while solving $\text{SOP}_\varsigma$ in~\eqref{SOP-1-n} and} compute the coefficients of the $\delta$-ISS Lyapunov functions along with other decision variables as
{\begin{align*}
		&\mathcal{S}_i^*(q_i,x_i,x_i')=~(x_{1_i}\!-\!x_{1_i}')^2 + 1.5(x_{2_i} - x_{2_i}')^2 ,\\
		&\mu_{\varsigma_{3_i}}^*\!\!= -0.9929,\mu_{\varsigma_{4_i}}^*\!\!=0.0002, \rho^*_i \!=\!10^{-7}, \underline{\alpha}_i^* \!=\! 1, \overline{\alpha}_i^* \!=\! 1.5,\\
		&\hat L_i^*=5, \bar q_i^{1*} = 1, \bar q_i^{2*} = 10^{-6}, \bar q_i^{3*} = 1.5,
	\end{align*}
	with a fixed $\gamma^*_i = 0.999$. 
	We compute $\varepsilon_{i}=0.1795$ according to Step~2 of Algorithm~\ref{Alg:2} using~\eqref{EQ:41}. We also compute $\mathscr L_i^{\mathcal S} = 4.2426$, $\mathscr L_i^1 = 1.4142$, and $\mathscr L_i^2 = 1.4142$ \emph{analytically}, as described in Remark~\ref{rem:lip analytic}. Given that the measured one-step transition is corrupted by noise with bound $\mathsf d = 0.01$ on its norm, {we assume $\mathscr L_i^{\hat f} = 0.06$ is given}. Given $\mathscr{L}_{i}^3 = 4.2774$, with $L_i=3$, computed using~\eqref{eq:Lq}, and $\max_{\bar{\mathcal N}_i}\|\hat f_i(\hat{x}_i^{zz'}\!\!,\!\hat{w}_i^{zz'}\!) \!-\! \hat f_i(\hat{x}_i^{z'\!z}\!,\!\hat{w}_i^{z'\!z})\| = 0.0468$, we have
	\begin{align*}
		\mu_{\varsigma_{3_i}}^* + \mathscr L_i^3\varepsilon_i = -0.2251\leq 0,
\end{align*}}
\!\!for all $i\in\{1,\dots,10000\}$, as per Theorem~\ref{Thm2}, and since the compositionality condition \eqref{Eq:43} is also fulfilled with {$\zeta = -9.999\times 10^{-4}$}, one can verify that the unknown interconnected dt-NS $\Sigma=\mathcal{I}(\Sigma_1,\dots,\Sigma_{10000})$ is $\delta$-GAS with respect to $x = 0$. Moreover, the incremental Lyapunov function {$\mathcal V(q, x,x') = \sum_{i=1}^{10000}\mathcal S_i^*(q_i,x_i,x_i') = \sum_{i=1}^{10000} \!((x_{1_i}\!-\!x_{1_i}')^2+ 1.5(x_{2_i} - x_{2_i}')^2)$} is valid for the interconnected dt-NS with a correctness guarantee. {Since all subsystems are identical, the subsystem-level SOP was solved once and the resulting certificate was used for all subsystems.} The computation took around {$34$} seconds on a MacBook Pro (Apple M2 Pro chip with 16GB memory).
\begin{figure}[t!]
	\centering
	\subfloat[{Data-driven incremental Lyapunov function of the interconnected network over time} \label{fig:V Net}]{
		\includegraphics[width=0.8\linewidth]{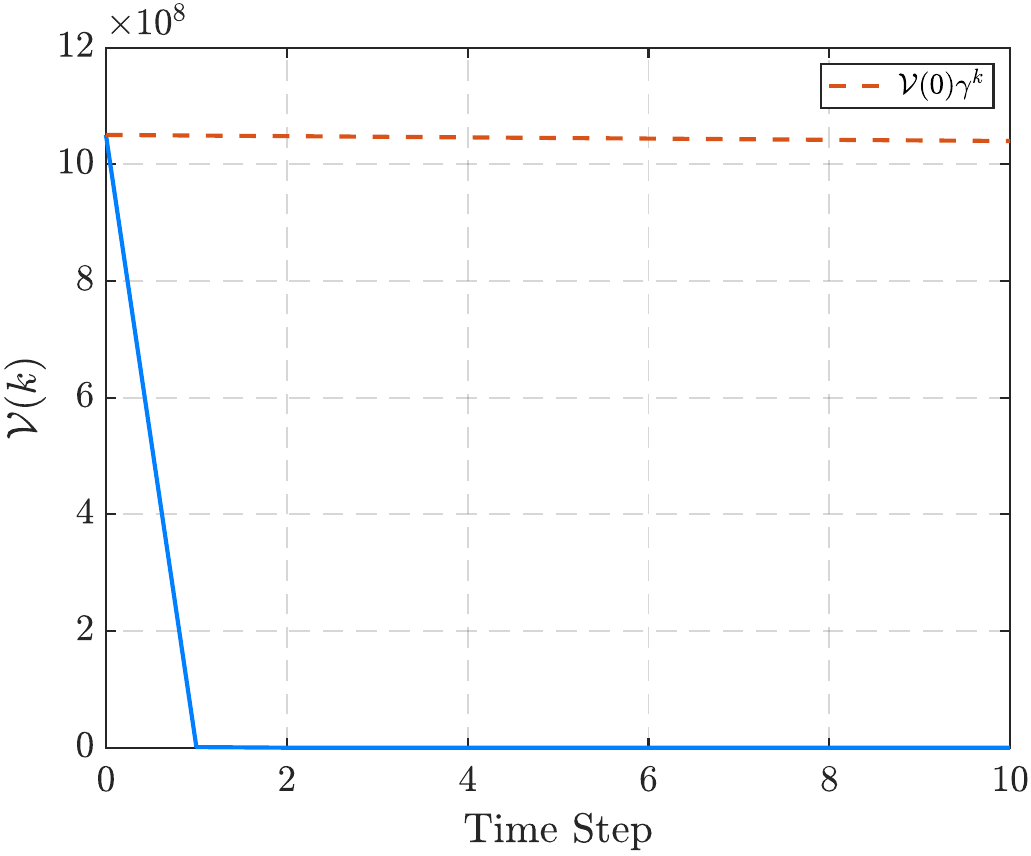}}
	\\
	\subfloat[{Sublevel sets of data-driven $\delta$-ISS Lyapunov function of an arbitrary subsystem}\label{fig:sublevel}]{
		\includegraphics[width=0.8\linewidth]{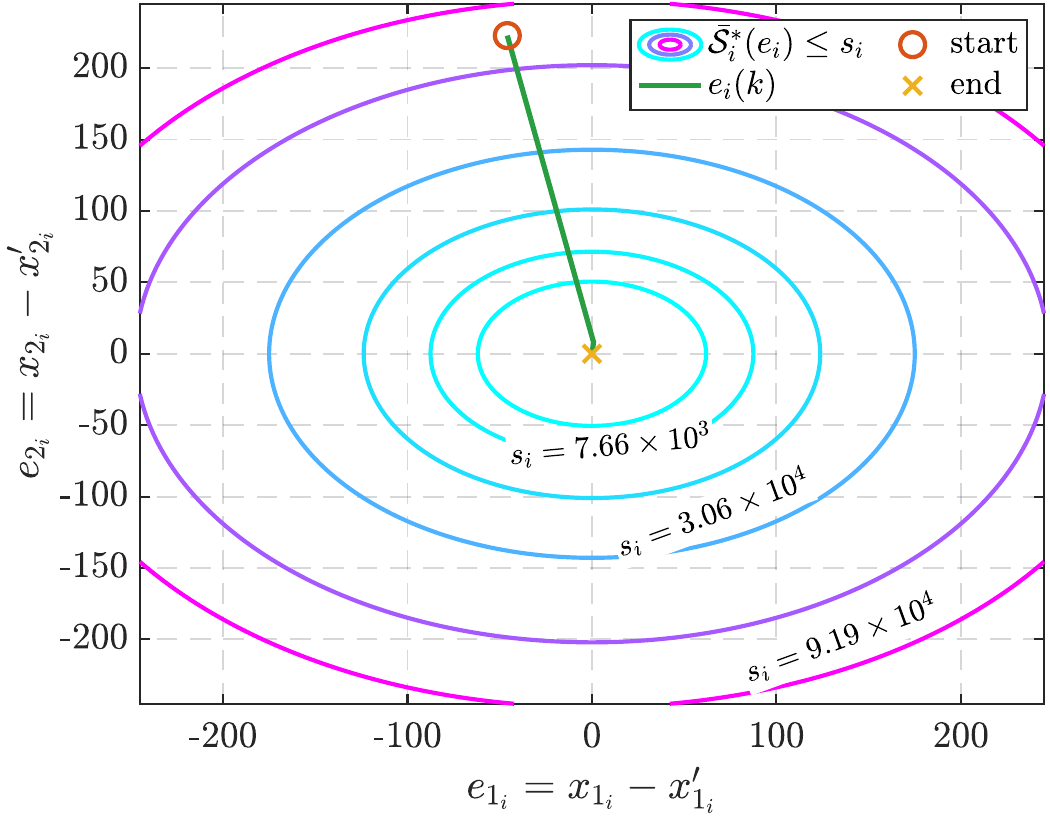}}
	\caption{{{\bf Incremental stability certificates:} (a) The incremental Lyapunov function of the controlled Duffing oscillator interconnected system decreases over time, verifying the satisfaction of condition~\eqref{alpha1}. (b) Illustration of the sublevel sets of the $\delta$-ISS Lyapunov function $\bar{\mathcal{S}}_i^*(e_i)\coloneq \mathcal S^*_i(q_i,x_i,x'_i)$ associated with an arbitrary subsystem, and the corresponding error-state trajectory $e_i\coloneq x_i - x'_i$. The initial and terminal states are highlighted, showing the contraction of the error dynamics toward the origin through nested sublevel sets of $\bar{\mathcal{S}}_i^*$.}}
	\label{fig:New}
\end{figure}
{\begin{remark}
		The use of identical subsystem dynamics is adopted solely for clarity of presentation and computational convenience, and does not restrict the generality of the framework, which is also applicable to heterogeneous interconnected systems.\hfill $\square$
\end{remark}}

Trajectories of the interconnected network initialized with arbitrary initial conditions $x(0)\in[-250~~250]^{20000}$ are illustrated in Fig.~\ref{fig:nonlinear}, showing the $\delta$-GAS property of the network. {Fig.~\ref{fig:nonlinear}(a) demonstrates that the trajectories $x_1$ and $x_1'$ obtained from one arbitrary subsystem starting from two different initial conditions converge to one another before converging to the origin asymptotically,} while Fig.~\ref{fig:nonlinear}(b) shows the corresponding trajectories $x_2$ and $x_2'$.

{Additionally, Fig.~\ref{fig:New} depicts incremental stability certificates associated with the interconnected network and one arbitrary subsystem. More precisely, Fig.~\ref{fig:New}(a) verifies that the incremental Lyapunov function obtained via our data-driven framework is strictly decreasing, satisfying condition~\eqref{alpha1}. In addition, Fig.~\ref{fig:New}(b) depicts the sublevel sets of the data-driven $\delta$-ISS Lyapunov function $\mathcal{S}_i^*(q_i, x_i, x'_i)$ for an arbitrary subsystem, together with the associated error trajectory $e_i(k)\coloneq x_i(k) - x'_i(k)$, indicating that the error dynamics evolve toward the origin while passing through nested sublevel sets of $\mathcal{S}_i^*$.}

\subsection{Sample Complexity Analysis}
\begin{figure} 
	\begin{center}
		\includegraphics[width=0.9\linewidth]{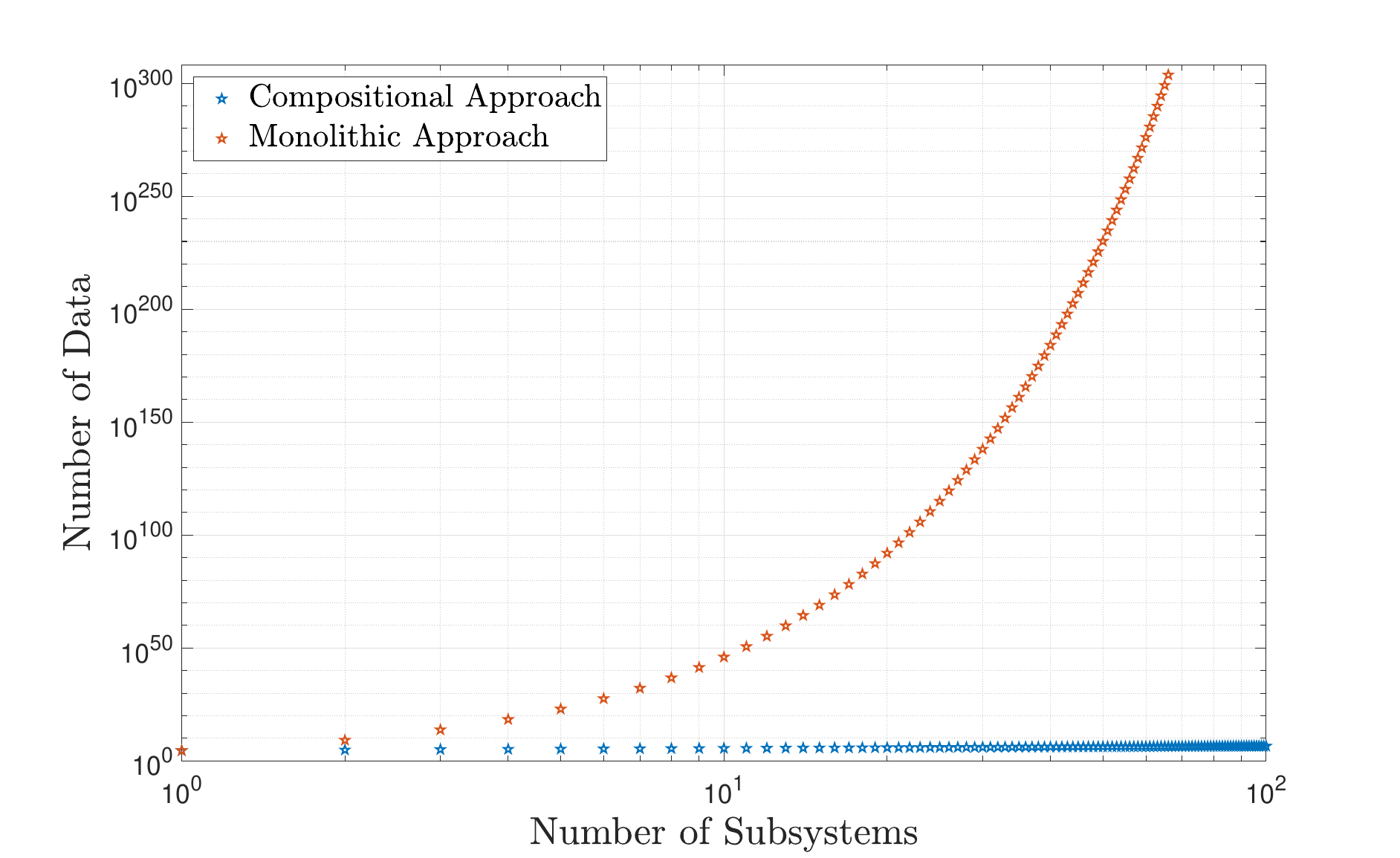} 
		\caption{{\bf Sample complexity comparison:} Required data versus subsystem number in monolithic and compositional methodologies on a logarithmic scale. {Our data-driven compositional framework restricts the exponential complexity to the dimension of each individual subsystem, rather than that of the overall interconnected network. By contrast, a monolithic treatment scales exponentially with the total network dimension, and thus rapidly becomes computationally intractable.}} 
		\label{Fig2}
	\end{center}
\end{figure}

Here, we conduct an examination of sample complexity within the contexts of both monolithic and compositional methodologies. The relationship between the required data points and the number of subsystems is visually depicted in Fig.~\ref{Fig2}. {At the subsystem level, the sample complexity of the proposed approach grows exponentially with twice the dimension of the subsystem state and its internal inputs. However, the key advantage of our data-driven compositional framework is that it confines this exponential dependence to the subsystem dimension, rather than the dimension of the entire interconnected network. Consequently, the overall sample complexity reduces to the subsystem level, since the SOP is solved per subsystem. In contrast, a monolithic approach exhibits exponential growth with respect to the total network dimension, quickly becoming computationally prohibitive.}

\section{Conclusion}\label{Discussion}
In this work, we proposed a data-driven \emph{compositional} approach to verify the incremental stability of nonlinear homogeneous networks of degree one, composed of subsystems with unknown dynamics. The proposed framework utilized incremental ISS properties of subsystems, characterized by the notion of $\delta$-ISS Lyapunov functions. In our data-driven setting, we collected {noisy} data from trajectories of each unknown subsystem and proposed a scenario optimization program (SOP) to enforce the necessary conditions for $\delta$-ISS Lyapunov functions. By solving the SOP proposed to incorporate noisy measurements, using the collected data, we constructed $\delta$-ISS Lyapunov functions for unknown subsystems. Subsequently, we built an incremental Lyapunov function for the interconnected system based on $\delta$-ISS Lyapunov functions of individual subsystems, by leveraging a compositionality condition derived from small-gain reasoning. To demonstrate the effectiveness of our data-driven approach, we applied it to a {physical} nonlinear homogeneous network consisting of $10000$ subsystems with unknown dynamics. The development of a data-driven method for designing incremental \emph{control} Lyapunov functions for large-scale networks with unknown dynamics is a subject of ongoing investigation for future work.

{\section*{Acknowledgment}\vspace{-0.2cm}
The authors would like to thank Behrad Samari for his assistance with the simulations in Section~\ref{Case_Study}.}
\bibliographystyle{agsm}
\bibliography{biblio}

\begin{authorbio}[Mahdieh]{Mahdieh Zaker} received her B.Sc. from K. N. Toosi University of Technology, Tehran, Iran, in 2019, and her M.Sc. from Amirkabir University of Technology (Tehran Polytechnic), Tehran, Iran, both in Electrical Engineering, control major. She is currently a PhD student in the School of Computing at Newcastle University, UK. She is the Best Repeatability Prize Finalist at the $28^{\text{th}}$ ACM International Conference on Hybrid Systems: Computation and Control (HSCC), 2025. Her research interests are (nonlinear) control and systems theory, data-driven techniques, large-scale systems, and formal methods.\vspace{0.7cm}
\end{authorbio}

\newpage

\begin{authorbio}[David]{David Angeli} is a
		Professor of Nonlinear Networks Dynamics within the Dept of Electrical and Electronic Engineering of Imperial College London.  
		He received the B.S. degree in computer science engineering and the PhD degree in control theory from the University of Florence, Florence, Italy, in 1996 and 2000, respectively.
		Since 2000, he has been an Assistant Professor and since 2005, an Associate Professor with the Department of Information Engineering, University of Florence. In 2007, he was a Visiting Professor with I.N.R.I.A de Rocquencourt, Paris, France, and in 2008, he joined as a Senior Lecturer the Department of Electrical and Electronic Engineering, Imperial College London, London, U.K., where he is currently a Professor and the Director of Postgraduate Teaching. He is the author of more than 120 journal papers in the research areas of stability of nonlinear systems, control of constrained systems (MPC), chemical reaction networks theory, and smart grids.
		Prof Angeli was an Associate Editor for the IEEE Transactions in Automatic Control and Automatica and was elevated to Fellow of the IEEE in 2015 for contributions to nonlinear control theory. He is a Fellow of the IET since 2018 and the recipient of the Honeywell Medal from InstMC in 2021.
\end{authorbio}

\begin{authorbio}[Abolfazl]{Abolfazl Lavaei} is an Assistant Professor in the School of Computing at Newcastle University, United Kingdom. Between January 2021 and July 2022, he was a Postdoctoral Associate in the Institute for Dynamic Systems and Control at ETH Zurich, Switzerland. He was also a Postdoctoral Researcher in the Department of Computer Science at LMU Munich, Germany, between November 2019 and January 2021. He received the Ph.D. degree in Electrical Engineering from the Technical University of Munich (TUM), Germany, in 2019. He obtained the M.Sc. degree in Aerospace Engineering with specialization in Flight Dynamics and Control from the University of Tehran (UT), Iran, in 2014. He is the recipient of several international awards in the acknowledgment of his work including  Best Repeatability Prize (Finalist) at the ACM HSCC 2025, IFAC ADHS 2024, and IFAC ADHS 2021, HSCC Best Demo/Poster Awards 2022 and 2020, IFAC Young Author Award Finalist 2019, and Best Graduate Student Award 2014 at University of Tehran with the full GPA (20/20). His research interests revolve around the intersection of Control Theory, Formal Methods in Computer Science, and Statistical Learning Theory.
\end{authorbio}

\end{document}